\documentclass[a4paper,11pt]{article}
\pdfoutput=1 % if your are submitting a pdflatex (i.e. if you have
\usepackage{jheppub} % for details on the use of the package, please
\usepackage{hyperref}
\usepackage[T1]{fontenc} % if needed

\title{\begin{center}
		\LARGE{\textbf{Seeing Higher-Dimensional Grand Unification \\In Primordial Non-Gaussianities}}
	\end{center}
}
\author{Soubhik Kumar}
\author{and Raman Sundrum}
\affiliation{Maryland Center for Fundamental Physics, Department of Physics,\\University of Maryland, College Park, MD 20742}
\emailAdd{soubhik@terpmail.umd.edu}
\emailAdd{raman@umd.edu}
\abstract{
The observed low-energy values of the $SU(3) \times SU(2) \times U(1)$ gauge couplings, extrapolated via the minimal Standard Model Renormalization Group evolution, hint at the exciting possibility of a Grand Unified Theory (GUT) at $M_{\text{U}}\sim 10^{14}$ GeV---a scale, however, too high to probe directly via collider searches.
Fortunately, since the Hubble scale $H$ can be as high as $5\times 10^{13}\text{GeV}\sim M_{U}$ during the inflationary era, such GUT scale states can be cosmologically produced at that time and leave direct \textit{on-shell} signatures such as their masses and spins, via primordial non-Gaussianity (NG).  We explore this possibility in one of its simplest 
realizations given by the extra-dimensional framework of orbifold GUTs, in which proton decay can be straightforwardly suppressed to be within the stringent bounds. Here, along with the massive GUT states there must also be $H$-mass spin-2  Kaluza-Klein (KK) gravitons, collectively giving rise to striking NG signatures. 
In our set-up we localize the inflaton on one of the boundaries of an extra dimension. 
The inflationary vacuum energy can readily lead to formation of a horizon in the bulk, where the KK modes then form a \textit{continuum} above a mass gap of $\sim {\cal O}(H)$. We find that the optimal case for observable NG signals is when the extra dimension is stabilized close to the onset of this horizon,  ensuring a \textit{discrete} KK spectrum such that the lightest KK modes can be cosmologically produced without significant Boltzmann suppressions. Although we mostly focus on the case where there is no higher-dimensional cosmological constant, we also obtain considerable holographic insights from the AdS$_5$/CFT$_4$ correspondence when such a cosmological constant is included.
}

\begin{document}
	\hspace{30em} UMD-PP-018-09
	\maketitle
	\flushbottom
\section{Introduction}
\begin{figure}[h]
	\centering
	\includegraphics[width=0.5\linewidth]{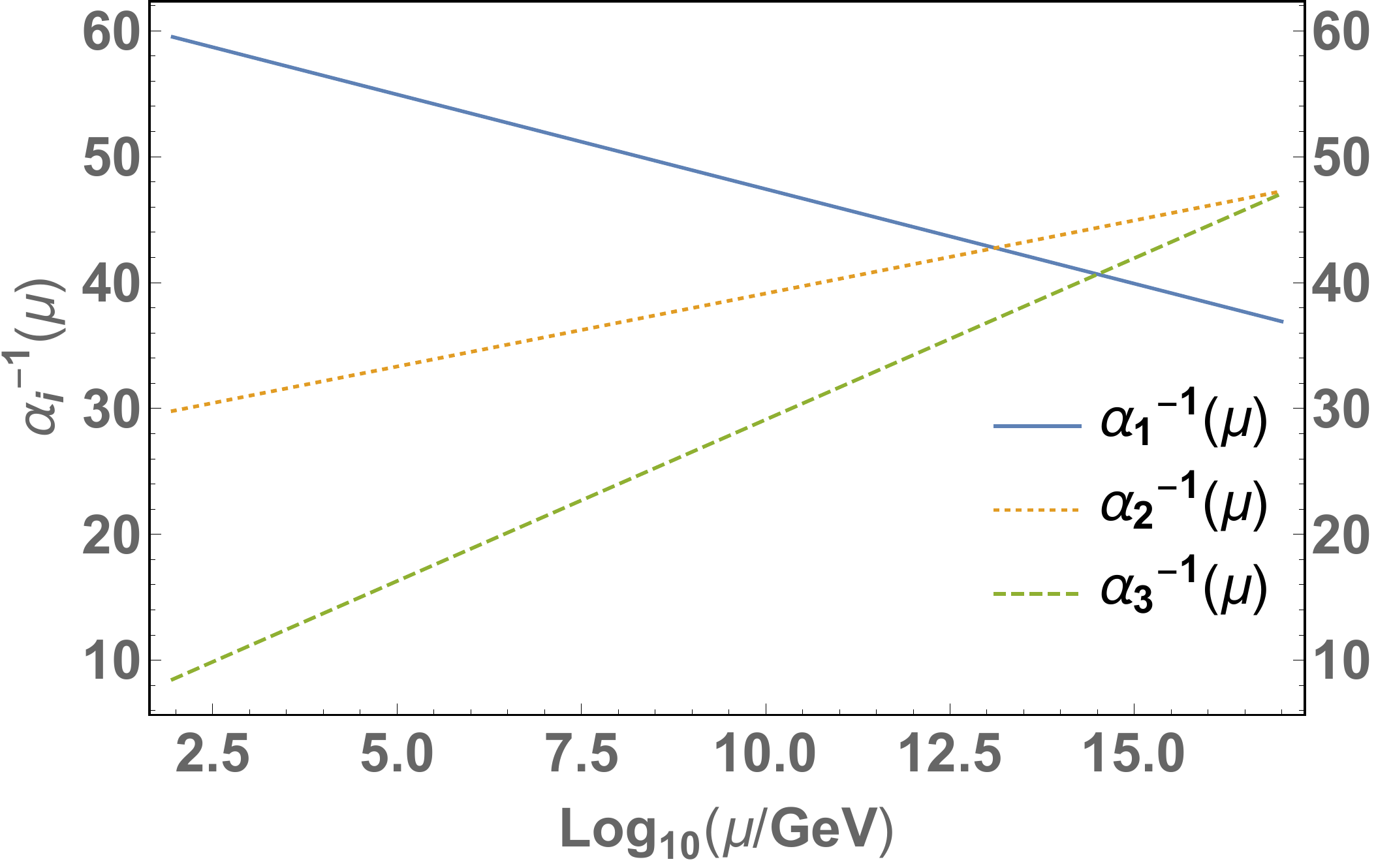}
	\caption{SM Renormalization Group Evolution (RGE) of gauge couplings $g_i$ at 1-loop written in terms of $\alpha_i \equiv g_i^2/4\pi$. The label ``i = 1,2,3'' denotes the $U(1)$ , $SU(2)$ and $SU(3)$ SM subgroups respectively with the normalization that $g_1 =􏰏\sqrt{5/3}g^\prime$ where $g^\prime$ is the SM hypercharge coupling.}
	\label{fig:thresold-correction-orbifold}
\end{figure}

It is an intriguing experimental fact that the $SU(3)\times SU(2)\times U(1)$ gauge couplings, when extrapolated using the minimal Standard Model (SM) Renormalization Group Evolution (RGE), become approximately equal to each other at an energy scale $M_U\sim 10^{14}$ GeV as seen from Fig. \ref{fig:thresold-correction-orbifold}.
This can be thought of as a strong circumstantial evidence for the attractive possibility that the SM gauge theory becomes part of a Grand Unified Theory (GUT) (see \cite{Tanabashi:2018oca} for a review) at that scale, characterized by a simple gauge group and a single gauge coupling. Some imperfection in the meeting of couplings at $M_U$, such as is seen in Fig. \ref{fig:thresold-correction-orbifold}, is to be expected from beyond-SM thresholds, either $\gtrsim$ TeV  as in the weak scale supersymmetric (SUSY) paradigm, or from splittings $\sim M_U$. In this paper, we consider the minimal scenario where only the non-supersymmetric SM exists in the infrared, with only $M_U$-scale threshold corrections from beyond the SM (BSM).

\begin{figure}[h]
	\centering
	\includegraphics[width=0.6\linewidth]{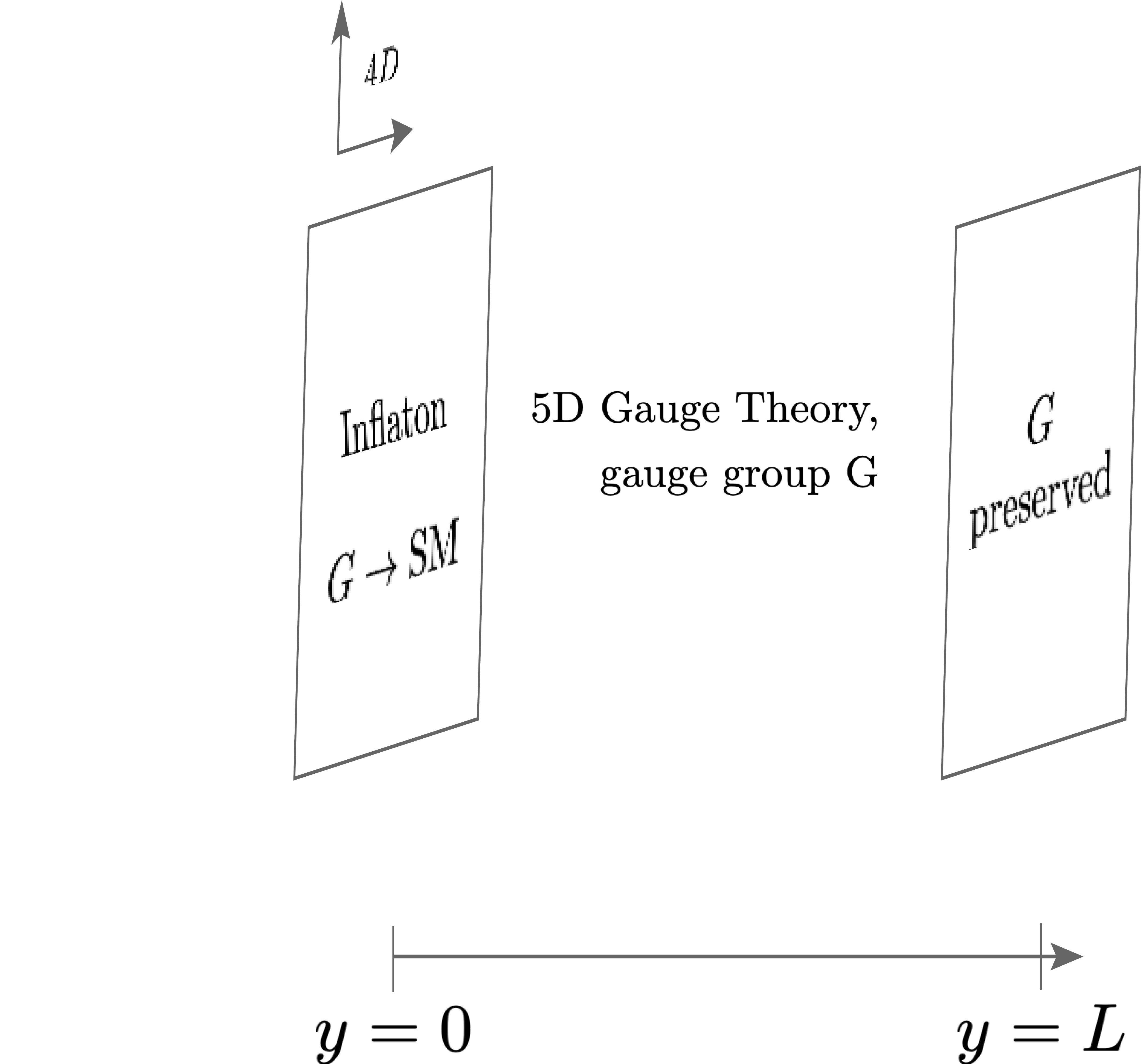}
	\caption{5D spacetime having two boundaries at $y=0$ and $y=L$. (a) Dirichlet Boundary Conditions (BC's) on the gauge bosons of GUT/SM achieves the breaking $G\rightarrow \text{SM}$ on the left boundary, also housing the inflaton $\phi(x)$. Neumann BC's on all gauge bosons preserve $G$ on the right boundary.}
	\label{fig:extra-dim-geometry}
\end{figure}

However, indirect constraints on such theories  exist \cite{Miura:2016krn}. 
In the simplest GUT gauge theories such as $SU(5)$ and $SO(10)$, unified matter multiplets contain both quarks and leptons,   leading to the prediction of proton decay mediated by GUT bosons.
Non-observation of proton decay then puts a lower bound, $M_U \gtrsim 10^{15}$ GeV, apparently ruling out minimal SM unification. 
While it is possible to build purely 4D models (for 
e.g. see the review \cite{Nath:2006ut} and references therein) that evade these stringent bounds, these are somewhat intricate. On the other hand,  
the extra dimensional framework of orbifold GUTs (see \cite{Kawamura:1999nj,Kawamura:2000ev,Hall:2002ea})  offers a very simple and plausible mechanism to suppress proton decay and still achieve unification. (Also see \cite{Weiner:2001pv,Csaki:2001qm} for orbifold GUT inspired 4D realizations.) In their simplest incarnations, orbifold GUTs are theories where a unified gauge theory lives in a (4+1)D spacetime with the extra dimension being an interval. Boundary conditions (BC's) on the bulk gauge fields then must be specified at the two ends of the interval and it is these conditions that determine which gauge fields will have zero modes and thus be present in the low energy theory. Since BC's need not respect the complete GUT gauge invariance, a breaking $\text{GUT}\rightarrow SU(3)\times SU(2)\times U(1)$ can be achieved simply through a suitable choice of BC's. See Fig. \ref{fig:extra-dim-geometry}.
The unification will only be manifest when we reach energy scales $\sim M_U\sim M_{KK}$, the mass of the lightest Kaluza-Klein (KK) excitations, that is at energies high enough to directly detect the extra dimension.
The proton decay bounds can be avoided by having separate GUT multiplets for SM quarks and leptons so that conserved baryon and lepton numbers can be consistently assigned to these multiplets \cite{Hall:2001pg}. 
Again, suitable boundary conditions on these 5D fermion multiplets can be imposed such that only the SM fermions have chiral zero modes and appear in the low energy effective theory. 

Without new TeV scale particles such as in SUSY or a robust proton decay signal, it seems impossible to directly test the orbifold GUT hypothesis at foreseeable colliders or other terrestrial experiments given that the non-SM states reside at $\sim M_U \sim M_{KK} \sim 10^{14}$ GeV. 
However, the primordial universe presents us with a unique opportunity in this regard. The Hubble scale $H$ during an era of cosmic inflation (see \cite{Baumann:2009ds} for a review) in the early universe could be as large as $5\times 10^{13}$ GeV \cite{Akrami:2018odb}, and hence GUT scale states having masses $M_U\sim H$ can be cosmologically produced during that era due to the time-dependence of the inflationary background. Furthermore, provided there is a suitable coupling, these states can decay into inflatons. This
can, in turn, give a very distinctive non-Gaussian contribution to the spectrum of primordial curvature fluctuations $\mathcal{R}$ \cite{Chen:2009zp,Baumann:2011nk,Assassi:2012zq,Chen:2012ge,Noumi:2012vr,Arkani-Hamed:2015bza,Dimastrogiovanni:2015pla,Lee:2016vti,Kehagias:2017cym,An:2017hlx,Kumar:2017ecc,Baumann:2017jvh,Franciolini:2017ktv,Arkani-Hamed:2018kmz},  that we can probe via the Cosmic Microwave Background \cite{Bartolo:2017sbu,Franciolini:2018eno}, Large-Scale Structure \cite{Schmidt:2015xka,Chisari:2016xki,Gleyzes:2016tdh,MoradinezhadDizgah:2017szk,MoradinezhadDizgah:2018ssw,MoradinezhadDizgah:2018pfo,Cabass:2018roz,Kogai:2018nse}, and more futuristically 21-cm cosmology \cite{Meerburg:2016zdz}. For various interesting applications of this idea, see the recent works e.g. \cite{Chen:2016nrs,Chen:2016uwp,Chen:2016hrz,Biagetti:2017viz,An:2017rwo,Kumar:2017ecc,Wang:2018tbf,Chen:2018xck,An:2018tcq,Chen:2018cgg}. Refs. \cite{Kumar:2017ecc,Chen:2016nrs,Chen:2016uwp,Chen:2016hrz,Chen:2018xck} discussed visibility (in the sense we describe now) of (B)SM Higgs, (B)SM gauge bosons and (B)SM fermions via primordial non-Gaussianity (NG).

\begin{figure}[h]
	\centering
	\includegraphics[width=0.9\linewidth]{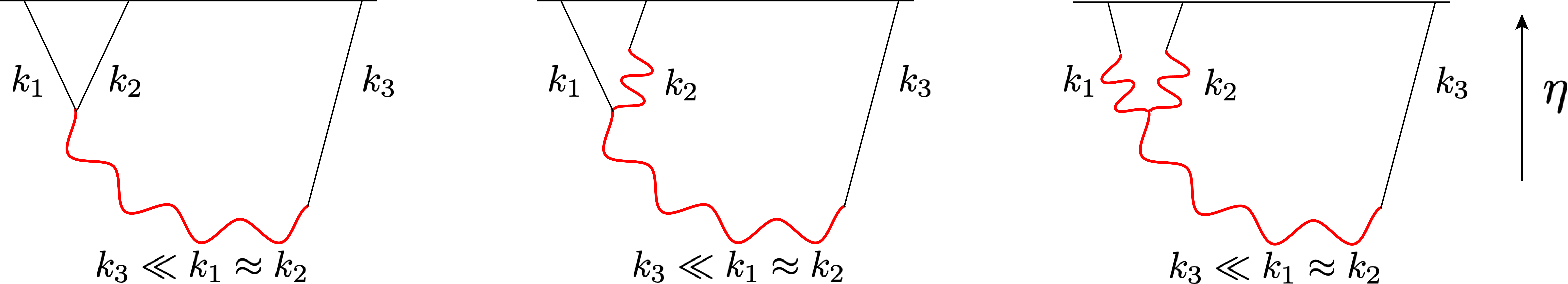}
	\caption{Tree level contributions to bispectrum due to massive particle exchange. From left to right: (a) single exchange diagram, (b) double exchange diagram, (c) triple exchange diagram. All the three diagrams depend on the mixing between the massive particle (in red) and the inflaton fluctuation (in black) in the (implicit) non-trivial background of slowly rolling $\phi_0(t)$. $\eta$ is (conformal) time, ending at the end of inflation.}
	\label{fig:diagrams}
\end{figure}

Let us briefly review the structure of these non-Gaussian contributions.
Massive fields with $H$-scale masses, if present during inflation with appreciable coupling to the inflaton, lead to a \textit{non-analytic} momentum dependence of the three-point function (i.e. the bispectrum) of $\mathcal{R}$ \cite{Chen:2009zp,Baumann:2011nk,Assassi:2012zq,Chen:2012ge,Noumi:2012vr,Arkani-Hamed:2015bza},
\begin{equation}\label{qsfi}
\langle \mathcal{R}(\vec{k}_1) \mathcal{R}(\vec{k}_2) \mathcal{R}(\vec{k}_3)\rangle \propto \mathcal{F}_s(\theta) \frac{1}{k_3^3}\frac{1}{k_1^3}\left(\frac{k_3}{k_1}\right)^{\Delta_s(m)}+\cdots, \text{ for } k_3\ll k_1,
\end{equation}
in the ``squeezed'' limit where one momentum is much smaller than the other two. 
Importantly, in eq. \eqref{qsfi}, the exponent $\Delta_s(m)$ and the pre-factor $\mathcal{F}_s(\theta)$, with $\theta = \vec{k}_3\cdot\vec{k}_1$, depend on the mass ($m$) and spin ($s$) of the massive particle. For example, for a spin-1 particle $\mathcal{F}_1(\theta)=\cos(\theta)$ and $\Delta_1(m)=\frac{5}{2}+i\sqrt{\frac{m^2}{H^2}-\frac{1}{4}}$ \cite{Lee:2016vti}. Thus a precise measurement of the bispectrum and its momentum dependence in the squeezed limit can capture the precious mass and spin information of the massive field. 
The contribution of such massive fields to the bispectrum can be represented by ``in-in'' diagrams, where the initial state is given approximately by the interacting Bunch-Davies de Sitter ``vacuum'' and the final time  is essentially the end of inflation. In particular we show in Fig. \ref{fig:diagrams} the three tree level contributions to the bispectrum which will be called single, double and triple exchange diagrams depending on the number of massive propagators.

The non-analytic momentum dependence in eq. \eqref{qsfi} signifies the fact that the massive particle is produced \textit{on-shell} during inflation and its effects can not be integrated out \cite{Arkani-Hamed:2015bza}. For $m\gg H$, the non-analytic contribution to the bispectrum will be very small since 
cosmological, on-shell productions of such heavy particles will be ``Boltzmann suppressed''. This suppression is captured by the proportionality factor in eq. \eqref{qsfi}, which we will write out explicitly in Secs. \ref{kkgravng} and \ref{kkgaugeng}. But when $m\ll H$ the non-analyticity in the three point function becomes insignificant as can be seen from the expression of $\Delta_1(m)$ above. Hence only the regime $m\sim H$ yields both a non-trivial and observable bispectrum carrying signatures of new physics. These are the primary observations behind the ambitious program of ``Cosmological Collider Physics'' \cite{Arkani-Hamed:2015bza} which has an unprecedented reach into the structure of fundamental physics at energy scales much higher than we can expect to probe at terrestrial colliders. 

Thus, the above considerations show that if during inflation $H$ is comparable to the GUT scale, then by studying primordial NG we may be able to do mass-spin spectroscopy  of GUT states! 
A robust feature of orbifold GUTs is that at the unification scale 
$\sim M_{KK}$ spacetime is necessarily higher-dimensional, and therefore there must be KK graviton excitations at this scale in addition to GUT/KK gauge states. 
This has two important, related consequences in the scenario we are focusing on with $H\sim M_{KK}$. First, the KK graviton will also have a mass $\sim H$ and a \textit{model independent} coupling to the inflaton, guaranteed by 5D diffeomorphism invariance. Therefore, in a set-up with orbifold GUTs, we expect to see not only the NG signatures of the GUT/KK gauge states but also striking spin-2 signatures due to KK gravitons. The second consequence is that, to describe inflationary dynamics completely, which involves energies $\sim H \sim M_{KK}$, we have to take into account the higher-dimensional geometry and cannot just  focus on a 4D effective theory where all the KK modes are integrated out.

The 5D geometry brings in a subtlety. To illustrate that, first consider a set-up where the inflaton is localized on one boundary of a {\it semi-infinite} extra dimension. The inflationary vacuum energy backreacts significantly on the 5D geometry and an event horizon will be formed at some finite distance, characterized by $H$, away from the inflationary boundary \cite{Vilenkin:1984hy,Ipser:1983db,Kaloper:1998sw,Kaloper:1999sm}. 
\begin{figure}[h]
	\centering
	\includegraphics[width=0.6\linewidth]{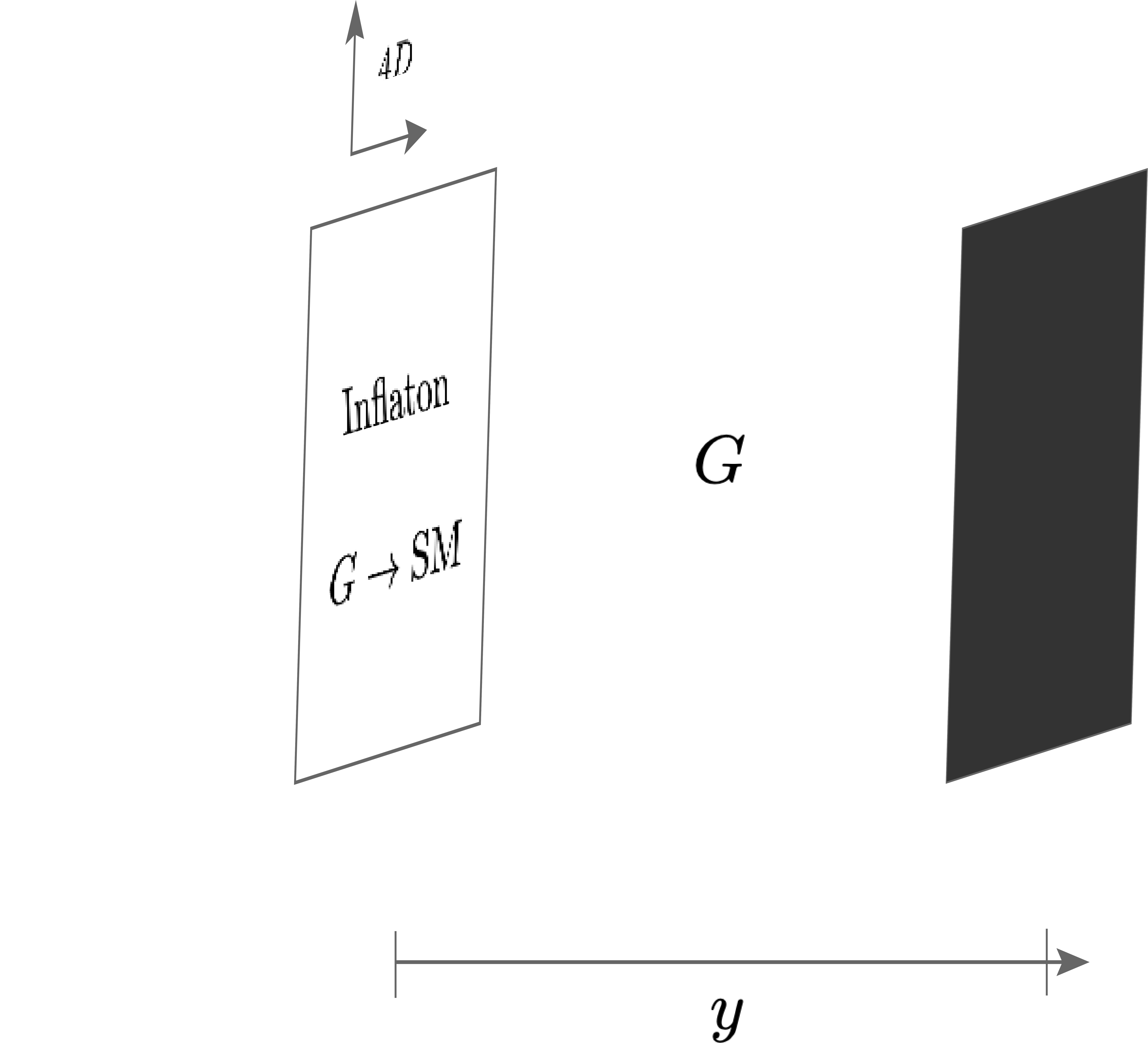}
	\caption{Same set-up as in Fig. \ref{fig:extra-dim-geometry} except the right boundary is absent and a ``black brane'' horizon has formed due to the backreaction of the inflationary vacuum energy on the left boundary.}
	\label{fig:extra-dim-geometry-horizon}
\end{figure}
See Fig. \ref{fig:extra-dim-geometry-horizon}.  Although such a horizon forms quite generally, it has a particularly nice holographic interpretation via the $\text{AdS}_5/\text{CFT}_4$ correspondence \cite{Maldacena:1997re} when there is a negative 5D Cosmological Constant (CC) in the bulk. The 5D spacetime is given by a detuned RS2 \cite{Randall:1999vf} set-up \cite{Kaloper:1999sm,Nihei:1999mt,Kim:1999ja},
dual to a purely 4D inflationary dynamics coupled to CFT$_4$ self-interacting radiation.
The temperature of the horizon, as we will show later, is equal to $\frac{H}{2\pi}$ which can be interpreted by the hot ``AdS/CFT'' correspondence (see e.g. \cite{Marolf:2010tg}) as the temperature of the dual 4D CFT. The CFT in this case is being heated due to the Gibbons-Hawking temperature \cite{Gibbons:1977mu} of $dS_4$. In this case, where the extra dimension is only cut off by a horizon, 
the KK spectra form a \textit{continuum} of states above a ${\cal O}(H)$  gap, dual to the states of the hot CFT plasma. 
On the other hand, we would like to do spectroscopy of a \textit{discrete} set of KK states, in a detuned RS1 set-up \cite{Randall:1999ee}, so we must ensure that the right boundary, in Fig. \ref{fig:extra-dim-geometry}, is stabilized to appear before the horizon is reached. The 4D dual statement is that the (deformed) CFT confines in the IR, but in order to do so the Gibbons-Hawking temperature must not exceed the deconfinement temperature. If this temperature is exceeded, the (deformed) CFT is deconfined, dual to the horizon in 5D appearing before the second boundary.  We will find that there is a ``window of opportunity'' for doing discrete spectroscopy using NG, constrained by the need for the de Sitter temperature $H/(2 \pi)$ to be below the deconfinement temperature, but not so low that cosmological production of the confined states (dual to discrete KK modes)  is Boltzmann suppressed. Studying this window will be a central part of our work. It is complicated by the fact that in this regime there is a significant backreaction on the Goldberger-Wise extra-dimensional stabilization mechanism \cite{Goldberger:1999uk} for the second boundary due to the $H$-scale inflation. We perform a novel near-horizon analysis in which this backreaction is systematically controllable. The final strength of NG signals will also depend on the backreaction away from the near-horizon regime, but only up to ${\cal O}(1)$ uncertainties, which do not affect their basic observability. We hope to address these uncertainties in later work. 

While the case of non-zero 5D CC offers a simple dual 4D interpretation, as above, we will mostly focus on the case of vanishing 5D CC for technical simplicity. However, the qualitative behavior is very similar to that with a CC, and the latter continues to provide good intuition for our results. Although our focus in this paper will be on the orbifold GUTs scenario, our results related to KK gravitons and stabilization of the extra dimension are quite general and will apply whenever the size of the extra dimension is ${\cal O}(H^{-1})$. 
 
This paper is organized as follows. In Sec. \ref{oguts} we detail the specific orbifold set-up that we will be considering in this paper and recall some aspects of the gauge coupling unification in orbifold GUTs. In particular, we will see that with boundary localized, non-GUT-symmetric 4D gauge kinetic terms one can easily fit the observed values of gauge couplings. In Sec. \ref{ininreview} we will briefly review the definitions of cosmological correlators and the ``in-in'' formalism used to calculate them. Sec. \ref{kkgrav} focuses on various extra dimensional features of the inflationary spacetime, as alluded to above, and ends with an estimation of the strength of NG mediated by KK gravitons. Sec. \ref{kkgauge} describes the inflationary couplings of the KK gauge bosons of the GUT, listing all the higher dimensional operators relevant for NG. We discuss the prospects of visibility when the GUT group is either $SU(5)$ or $SO(10)$. Secs. \ref{kkgravng} and \ref{kkgaugeng} give the explicit form of NG mediated by KK gravitons and KK gauge bosons respectively and calculate the strengths of NG. We conclude in Sec. \ref{conclusion}. Two technical appendices supplement the discussion in the main text of the paper. In Appendix \ref{appkk} we derive the KK decomposition of the KK graviton-radion system, both reproducing some of the existing results from the literature and establishing some new results that are used in Sec. \ref{kkgrav}. In Appendix \ref{appA} we derive the bispectrum mediated by KK gravitons via a direct computation using the ``in-in'' formalism. This, as it should, reproduces the form of \cite{Arkani-Hamed:2015bza} obtained via exploiting conformal symmetries of the  late time slice. Furthermore, our calculation also determines the overall normalization of the in-in correlator.

\section{Orbifold GUTs and Gauge Coupling Unification}\label{oguts}

\subsection{Orbifold GUTs}  We consider the simplest orbifold GUT structure with a 5D bulk, and with a simple GUT gauge group such as $SO(10)$ or $SU(5)$. The 5D gauge theory is necessarily a non-renormalizable effective field theory (EFT). The extra dimension is physically an interval, although we will realize this as an $S^1/(Z_2 \times Z_2')$ quotient of a circle in order to precisely specify boundary conditions. 
While the 5D bulk preserves the  GUT gauge symmetry, it is broken on one of the boundaries of the extra-dimensional interval (effectively Higgsed at the 5D EFT cutoff) down to just the SM gauge group. We can think of this as effectively being given by imposing Dirichlet boundary conditions (BC's) on the broken gauge fields and Neumann BC's on the unbroken (SM) gauge fields on the GUT-breaking boundary, and all-Neumann BC's on the other GUT-symmetric boundary. Lastly, we will take the SM fermions and the SM Higgs to be present in the bulk as well.

Before discussing gauge coupling unification in such a set-up, we give the explicit extra dimensional profiles of the KK modes of the bulk gauge bosons given our choice of BC's. In this section we will assume a simple fixed 5D spacetime product geometry consisting of 4D Minkowski spacetime and the extra-dimensional interval. We will account for 5D curvature in later sections, but this will not change the central structure of unification and its low-energy implications. For finding the free-field profiles we can ignore the self-interactions of the bulk non-Abelian gauge field. 

Then the equation of motion (EOM) for each gauge field component is identical to the Maxwell equations for a bulk $U(1)$ gauge field. These are given by (suppressing the adjoint index on the gauge field),
\begin{equation}
\partial_M F^{MN} = 0.
\end{equation}
By a suitable gauge transformation we can go to the gauge where $A_5(x,y)=A_5(x)$ with $y$ being the coordinate along the extra dimension. Furthermore, with our choice of BC's above, $A_5(x)=0$. Hence the Maxwell equations for $A_\nu$ are given by,
\begin{equation}\label{eomgauge1}
\partial^\nu F_{\nu\mu}+\partial^2_y A_\mu = 0.
\end{equation}
Via a KK decomposition,
\begin{equation}
A_\mu = \sum_{l}A_{l,\mu}(x)\vartheta_l(y),
\end{equation}
the 5D EOM \eqref{eomgauge1} can be separated into a 4D EOM for a massive spin-1 particle and an equation governing the extra dimensional profile,
\begin{eqnarray}
\partial^\nu F_{l,\nu\mu} = m_l^2A_{l,\mu}, \\ \label{kkgaugeprofile}
\partial_y^2 \vartheta_l+m_l^2 \vartheta_l = 0.
\end{eqnarray}
Here $m_l$ is the mass of the $l$-th KK mode. Using eq. \eqref{kkgaugeprofile} we can derive the profile of SM and broken gauge fields (part of GUT/SM coset) for the above choice of BC's,
\begin{eqnarray}\label{SMzeromode}
\vartheta_l^{\text{SM}}(y)=\cos(l\pi y/L),\\
\vartheta_l^{\text{GUT/SM}}(y)=\sin((l+1/2)\pi y/L),
\end{eqnarray}
with $l$ being a non-negative integer. %We have also put back the adjoint index on the profiles $\vartheta$.  
We have placed the boundaries at $y=0$ and $y=L$. Taking $l=0$ we see that  only the SM gauge bosons have a zero mode, $m=0$, whereas the lightest of GUT/SM bosons have a mass of $m = \frac{\pi}{2L}$ and hence no zero mode. This choice of BC's has broken the GUT down to the SM at the compactification scale, as expected.

\subsection{Gauge Coupling Unification} The action for the gauge sector is given by,
\begin{equation}\label{laggauge}
S\supset\int d^4 x \int_{0}^{L}dy \sqrt{-G} \left(\frac{1}{g_5^2}F_{MN}F^{MN}+\delta(y)\sum_{i}\kappa_i F_{i ,\mu\nu}F^{\mu\nu}_i\right),
\end{equation}
where $M,N$ and $\mu,\nu$ run over the 5D and 4D indices respectively. The first term describes the field strength for the bulk GUT gauge theory. For generality, we have also included boundary localized, \textit{non}-GUT-symmetric 4D gauge kinetic terms, where the label ``$i = 1,2,3$'' denotes the $U(1)$, $SU(2)$ and $SU(3)$ SM subgroups. We can now relate the gauge couplings $g_{4,i}$ \footnote{Here we are making a small change in notation compared to Fig. \ref{fig:thresold-correction-orbifold} by making the replacement $g_i\rightarrow g_{4,i}$ for $i = 1,2,3$.} in the 4D low energy effective theory with the 5D gauge coupling $g_5$. To this end, we note that the zero modes of the gauge bosons have a flat profile, as seen from eq. \eqref{SMzeromode} for $l=0$, in the extra dimension. Then using the Lagrangian \eqref{laggauge} and doing an integration over the extra dimension we get the  relation between the SM gauge couplings at the compactification scale (see e.g. \cite{Hall:2002ea}),
\begin{equation}\label{unification}
\alpha_i^{-1}\left(\frac{1}{L}\right)\equiv\frac{4\pi}{g_{4,i}^2}=\frac{4\pi L }{g_5^2}+ 4\pi \kappa_i.
\end{equation}
Below the unification scale $m_{KK} \sim 1/L$, the couplings $g_{4,i}$ evolve as per the usual SM RGE which at 1-loop reads as,
\begin{eqnarray}\label{running-gut}
\alpha_i^{-1}(\mu)=\frac{4\pi L }{g_5^2} + \frac{b_i}{2\pi}\log\left( \frac{m_{KK}}{\mu}   \right)+4\pi \kappa_i. 
\end{eqnarray}
In the above $b_i=(\frac{41}{10},-\frac{19}{6},-7)$ are the three 1-loop SM beta functions with the notation that $g_{4,1}=\sqrt{5/3}g'$ where $g'$ is the SM hypercharge coupling. This has precisely the one-loop form of a traditional 4D GUT, if we translate $\alpha_{GUT} = g_5^2/4 \pi L, M_{GUT} = m_{KK}$ and the $\kappa_i$ are interpreted as GUT threshold corrections.  We see that for sufficiently large $L$ and sufficiently long running, the first two terms on the right dominate, with the ``threshold corrections'' $\kappa_i$ giving a subleading contribution. This structure then predicts that plotting $1/\alpha_i(\mu)$ vs. $\log \mu$ will give three lines almost meeting at a point, as indeed the data suggests in Fig. \ref{fig:thresold-correction-orbifold}.\footnote{The minimal radiatively stable size of the $\kappa_i$ is $\sim \frac{1}{16 \pi^2}$. But it is perfectly natural for the the $\kappa_i$ to take larger values, required to interpret Fig. \ref{fig:thresold-correction-orbifold} in the orbifold GUT scenario as we do here.}
 As a benchmark choice taking, $\kappa_1= \frac{40}{16\pi^2}, \kappa_2= \frac{60}{16\pi^2}$ and $ \kappa_3= \frac{1}{16\pi^2}$ we can describe the observed gauge couplings at the weak scale and achieve unification in the sense described above with,
\begin{equation}
\alpha_{G}^{-1} = 39,\hspace{3em} m_{KK}= 5\times 10^{13} \text{GeV}.
\end{equation}

The lower unification scale of non-supersymmetric GUTs $\sim 10^{14}$ GeV raises the danger of an unacceptably large proton decay rate mediated by GUT states. In orbifold GUTs this is straightforwardly avoided by the mechanism of ``split multiplets'' whereby SM quarks and leptons are housed within different GUT multiplets, so that baryon and lepton number can be separately assigned, and unwanted fermionic zero-modes in these multiplets are removed by Dirichlet BC's on the GUT-breaking boundary \cite{Hall:2001pg}.

\section{Cosmological Correlators and Primordial Non-Gaussianity}\label{ininreview}
\subsection{Cosmological Correlators}
In this subsection we will very briefly summarize how cosmological correlation functions are defined and the formalism used to calculate them. For a more thorough explanation of this the reader is referred to our previous work \cite{Kumar:2017ecc}, along with the literature \cite{Weinberg:2005vy,Chen:2010xka}. 
\paragraph{Master Formula.} To calculate primordial non-Gaussianity (NG) due to inflaton fluctuations, ``in-in'' expectation values of some gauge invariant observable of interest are evaluated. This is most conveniently done in the interaction picture in which the master formula for ``in-in'' expectation value of a gauge invariant observable $Q$ at a time $t_f$ towards the end of inflation reads as,

	\begin{equation}\label{ininmasterformula}
 \langle 0\vert \bar{T} e^{+i\int\limits_{-\infty(1+i\epsilon)}^{t_f} dt_2 \mathsf{H}_I^{\text{int}}(t_2) } Q_I(t_f) T e^{-i\int\limits_{-\infty(1-i\epsilon)}^{t_f} dt_1 \mathsf{H}_I^{\text{int}}(t_1) }\vert 0\rangle.
	\end{equation}	
In the above, $\mathsf{H}_I^{\text{int}}$ is the interacting part of the full Hamiltonian, $\mathsf{H}=\mathsf{H}^{\text{quadratic}}+\mathsf{H}^{\text{int}}$ evaluated in the interaction picture. The operator $Q$ is denoted by $Q_I$ after being evaluated in the interaction picture. $\vert 0\rangle$ is the \textit{free} vacuum obtained from the \textit{interacting} vacuum by letting the early time evolution be along a slightly complex direction. $T(\bar{T})$ denotes time (anti-time) ordered product.

\paragraph{Choice of Gauge.}
Before going into the definition of various cosmological correlation functions let us address the issue of gauge invariance. We can split the inflaton field $\phi(t,\vec{x})$ into a homogeneous background field $\phi_0(t)$ and a fluctuation field $\xi(t,\vec{x})$, $\phi(t,\vec{x})=\phi_0(t)+\xi(t,\vec{x})$. 
Now, the scalar fluctuation $\xi$ mixes with scalar fluctuations coming from the metric and is not gauge invariant in general. A gauge invariant observable, which we will denote by $\mathcal{R}$, capturing scalar fluctuations can be constructed (see \cite{Weinberg:2008zzc} and references therein). However, rather than working with $\mathcal{R}$ ``in-in'' calculations can often be simplified by choosing the spatially flat gauge \cite{Maldacena:2002vr}. In this gauge the spatial part of the metric contains just the transverse traceless tensor $\gamma_{ij}$,
\begin{equation}\label{spatiallyflat}
h_{ij} = a^2(t)\left(\delta_{ij}+\gamma_{ij}\right),
\end{equation}
and up to slow-roll corrections we can treat $\xi(t,\vec{x})$ as a massless field in a fixed background inflationary spacetime metric \cite{Maldacena:2002vr} given by,
\begin{equation}
ds^2=-dt^2+ a^2(t)d\vec{x}^2.
\end{equation}
In the above, $a(t)=e^{Ht}$ is the scale factor in terms of Hubble scale $H$, which is a constant to the leading order in $\dot{\phi}_0,\ddot{\phi}_0$. Finally, $\mathcal{R}$ is related to $\xi(t,\vec{x})$ by
\begin{equation}\label{Rtoxi}
\mathcal{R}=-\frac{H}{\dot{\phi}_0}\xi.
\end{equation}
We will calculate ``in-in'' results in terms of $\xi$ and then use the above eq. \eqref{Rtoxi} to get a gauge invariant answer by rewriting the correlators in terms of $\mathcal{R}$.

\paragraph{Power Spectrum and Higher Point Functions.}
An n-point correlation function of scalar fluctuation $\mathcal{R}$ can be defined as the ``in-in'' expectation value
$\langle\mathcal{R}(\vec{k}_1) \mathcal{R}(\vec{k}_2)\cdots\mathcal{R}(\vec{k}_n)\rangle $. Note since we calculate the expectation value at a fixed instant of time, it is only the three momenta that appear in the above expression. It is conventional to strip momentum conserving delta functions and define,
	\begin{equation}
	\langle\mathcal{R}(\vec{k}_1)\cdots\mathcal{R}(\vec{k}_n)\rangle = (2\pi)^3\delta^3(\vec{k}_1+\cdots+\vec{k}_n)\langle\mathcal{R}(\vec{k}_1)\cdots\mathcal{R}(\vec{k}_n)\rangle^\prime.
	\end{equation}
The scalar power spectrum $P_{S,k}$ can then be calculated as
\begin{equation}
	P_{S,k}\equiv\langle\mathcal{R}(\vec{k})\mathcal{R}(-\vec{k})\rangle^\prime = \frac{H^4}{\dot{\phi}_0^2}\frac{1}{2k^3},
\end{equation}
where the r.h.s is to be evaluated at the time of horizon exit $k=aH$ for a given $k-$mode. Planck data \cite{Akrami:2018odb} gives the magnitude and tilt of the power spectrum to be $\frac{H^4}{\dot{\phi}_0^2}\approx 8.2\times 10^{-8}$ and $n_s\approx 0.96$ at a ``pivot'' scale 0.05 $\text{Mpc}^{-1}$.

The three point function i.e. the bispectrum can be defined in a similar way,
	\begin{equation}\label{Bfunction}
	B(k_1,k_2,k_3)\equiv	\langle \mathcal{R}({\vec{k}_1})\mathcal{R}({\vec{k}_2})\mathcal{R}({\vec{k}_3})\rangle^\prime.
	\end{equation}
A dimensionless bispectrum that we will often use in the rest of paper can be defined by,	
\begin{equation}\label{Ffunction}
F(k_1,k_2,k_3)\equiv\frac{B(k_1,k_2,k_3)}{P_{S,k_1}P_{S,k_3}}.
%\vert_{k_3\ll k_1,k_2}.
% \equiv f_{\text{NL}}\times f(k_1,k_2,k_3)
\end{equation}
In the spatially flat gauge using eq. \eqref{Rtoxi} this can be rewritten as
\begin{equation}\label{F}
F(k_1,k_2,k_3)=-\frac{\dot{\phi}_0}{H}\frac{\langle\xi({\vec{k}_1})\xi({\vec{k}_2})\xi({\vec{k}_3})\rangle'}{\langle\xi({\vec{k}_1})\xi({-\vec{k}_1})\rangle' \langle\xi({\vec{k}_3})\xi({-\vec{k}_3})\rangle'}.
\end{equation}
The size of NG is typically quoted in the literature as a constraint on a parameter $f_{\text{NL}}$ which is defined as
	\begin{equation}\label{fnl}
	f_{\text{NL}} \equiv \frac{5}{18} F(k,k,k).
	\end{equation}
In general the function $F(k_1,k_2,k_3)$ can have some non-trivial momentum dependent ``shapes''. CMB constraints on $f_{\text{NL}}$ depending on differing shapes can be found in \cite{Ade:2015ava} with a rough precision being $\sigma_{\text{f}_{\text{NL}}}\sim \mathcal{O}(5-50)$. This sensitivity is expected to be improved in future with LSS experiments to $\sigma_{\text{f}_{\text{NL}}}\sim \mathcal{O}(1)$ \cite{Alvarez:2014vva}. However, the ultimate sensitivity on this will come from an only cosmic variance limited 21-cm experiment because of the enormous number of modes, $N_{\text{21-cm}}$, such an ideal experiment can access. Very roughly we can have
\begin{equation}\label{cosmicvar}
\frac{\langle\mathcal{R}\mathcal{R}\mathcal{R}\rangle}{\langle\mathcal{R}\mathcal{R}\rangle^{\frac{3}{2}}}\sim \frac{1}{\sqrt{N_{\text{21-cm}}}}\sim 10^{-8},
\end{equation}
where with only cosmic variance, $N_{\text{21-cm}}$ can be as large as $10^{16}$ \cite{Loeb:2003ya}. This can help us achieve $\sigma_{\text{f}_{\text{NL}}}\sim \mathcal{O}(10^{-4}-10^{-3})$. This is the sensitivity that we will keep in mind when discussing the observability of our signals. We make the crucial assumption that \textit{non}-primordial NG induced by various non-linear effects after the modes re-enter the horizon can be modeled accurately enough so as to extract the primordial contribution.

\subsection{Non-gaussianity and Massive Particles}
Inflaton self-interactions or presence of other light fields with masses $\ll H$, can contribute to primordial NG (see \cite{Chen:2010xka} for a review and references to original papers). However, a very distinctive non-Gaussian feature of primordial fluctuations can emerge, if massive fields with $m\sim H$ are produced during inflation with sufficiently strong coupling to the inflaton, in the ``squeezed'' limit when one of the inflaton momenta becomes much smaller than the others (say, $k_3\ll k_1\sim k_2$). Depending on the mass ($m$) and spin ($s$) of such a particle, the bispectrum mediated by it will have a \textit{non-analytic} momentum dependent part of the form \cite{Chen:2009zp,Baumann:2011nk,Assassi:2012zq,Chen:2012ge,Noumi:2012vr,Arkani-Hamed:2015bza,Lee:2016vti},
\begin{eqnarray}\label{fnl_qsfi}
F^{\text{nonanalytic}}_{s=0}  \propto& f_0(\mu_0)\left(\frac{k_3}{k_1}\right)^{\frac{3}{2}+i\mu_0}+f_0(-\mu_0)\left(\frac{k_3}{k_1}\right)^{\frac{3}{2}-i\mu_0},\\
F^{\text{nonanalytic}}_{s=1}  \propto& \sin^2\theta \times \left( f_1(\mu_1)\left(\frac{k_3}{k_1}\right)^{\frac{5}{2}+i\mu_1}+f_1(-\mu_1)\left(\frac{k_3}{k_1}\right)^{\frac{5}{2}-i\mu_1}\right),\\\label{fnl_qsfi2}
F^{\text{nonanalytic}}_{s=2}  \propto& \left(\cos^2\theta-\frac{1}{3}\right) \times \left( f_2(\mu_2)\left(\frac{k_3}{k_1}\right)^{\frac{3}{2}+i\mu_2}+f_2(-\mu_2)\left(\frac{k_3}{k_1}\right)^{\frac{3}{2}-i\mu_2}\right).
\end{eqnarray}
In the above, $\mu_0=\mu_2=\sqrt{\frac{m^2}{H^2}-\frac{9}{4}}$ and $\mu_1=\sqrt{\frac{m^2}{H^2}-\frac{1}{4}}$ are given in terms of the mass $m$ of the massive particle. The spin dependence is encoded in the prefactors with $\theta=\hat{k}_1\cdot \hat{k}_3$. The non-analytic dependence on momenta also follows from simple considerations as reviewed in \cite{Kumar:2017ecc}. The functions $f_s(\mu_s)$ can be calculated given the coupling between inflaton and the massive particle. For the detailed form of $f_s(\mu_s)$ see e.g. \cite{Chen:2009zp,Baumann:2011nk,Assassi:2012zq,Chen:2012ge,Noumi:2012vr,Arkani-Hamed:2015bza,Lee:2016vti,Chen:2017ryl,Kumar:2017ecc} for spin-0; \cite{Kumar:2017ecc} for spin-1; eq. \eqref{f2} of the present paper and  \cite{Arkani-Hamed:2015bza,Lee:2016vti,Arkani-Hamed:2018kmz} for spin-2. While the Hubble spacetime expansion can readily produce particles with masses of order $H$ or smaller, for larger masses there is a ``Boltzmann suppressed'' production amplitude,  
generically  $f_s(\mu_s)\sim e^{-\pi\mu_s} \sim e^{-\pi m/H}$ for $m \gg H$. 
While extra dimensions certainly give rise to higher spin particles such as our KK gravitons, with a lower bound on their masses to avoid horizon formation, there is an even more robust bound on higher spin masses in 4D dS spacetime regardless of their origin. For spin-2 this is given by the Higuchi bound, \cite{Higuchi:1986py}. We will show that horizon non-formation is a stronger condition in the extra dimensional scenario so that the Higuchi bound is automatically satisfied.

Importantly, the non-analytic momentum dependence shown above cannot be ``faked'' by inflaton self-interactions since the NG contribution of the latter have only an analytic momentum dependence in the squeezed limit---making the non-analyticity a ``smoking gun'' signal of new particles during inflation \cite{Arkani-Hamed:2015bza}. Thus from a precision measurement of the bispectrum in the squeezed limit, we can probe particles with masses comparable to $H$ and their spins, far beyond the reach of terrestrial colliders---this is the goal of the ambitious ``Cosmological Collider Physics'' \cite{Arkani-Hamed:2015bza} program.

\section{Inflation and the Fifth Dimension}\label{kkgrav}
\subsection{General Set-up}
We consider a 5D spacetime in which the extra dimension is an interval and localize a 4D inflaton on one of the boundaries at an end of the interval. 
Technically, we will realize this interval as an $S^1/(Z_2\times Z_2^\prime)$ orbifold in order to determine BC's.
Set-ups with boundary localized inflaton have been considered in the literature, see e.g. \cite{Kaloper:1998sw,Lukas:1999yn,Kaloper:1999sm,Nihei:1999mt,Kim:1999ja,Maartens:1999hf,Giudice:2002vh,Im:2017eju}. %The bulk cosmological constant will be assumed to vanish, unless explicitly mentioned. 
%In this subsection we will lay out some general features of this set-up, which will be used in subsequently.
We will see that for sufficiently large $H$ the non-inflaton boundary can become shrouded by a black brane horizon, effectively leaving a set-up with a single boundary. To most simply explore this, we will also consider the limiting case of semi-infinite extra dimension.

The 5D action has the basic structure,
\begin{align}
S = \int d^4x \int_{0}^{L}dy \sqrt{-G}(2M_5^3R_5-\Lambda_5) - \int d^4 x \int_{0}^{L}dy \sqrt{-G}\delta(y) V_0 \nonumber\\ -\int d^4 x \int_{0}^{L}dy \sqrt{-G} \delta(y-L) V_L + 
\int d^4x \int_{0}^{L}dy \sqrt{-G} \left(-\frac{1}{2}G^{MN}\partial_M \Sigma\partial_N \Sigma - V(\Sigma)\right),
\end{align}
where the bulk metric is denoted by $G_{MN}$ and $G=\text{det}(G_{MN})$. $R_5$ is the 5D Ricci scalar. Here we have placed the boundaries at $y=0$ and $y=L$ where $y$ is the coordinate along the extra dimension. There are boundary-localized potentials $V_0,V_L$ at $y=0$ and $y=L$ respectively. $M_5$ is the 5D Planck scale whereas $\Lambda_5$ is the 5D cosmological constant.
We take the inflaton field to live at $y=0$, but begin by neglecting its rolling, so that its potential is a approximately constant $V_0 \sim M_4^2 H^2$, where $M_4$ is the final effective 4D Planck scale. We will however consider $V_{L}$ to be an exactly constant ``brane tension''.
We also have a bulk 5D Goldberger-Wise (GW) scalar $\Sigma$ \cite{Goldberger:1999uk} with a potential $V(\Sigma)$ that stabilizes the extra dimension. 
The case with a single boundary will be realized by taking  $L\rightarrow \infty$ limit.

Requiring a $dS_{4}$ foliation (in the limit of no-rolling of the inflaton) and a static extra dimension we are lead to the ansatz,
\begin{equation}\label{bgmetric}
ds^2= -n(y)^2 dt^2 + n(y)^2a(t)^2  d\vec{x}^2 + dy^2,
\end{equation}
where $a(t)=e^{ H t}$ is the scale factor, and $n(y)$ is the warp factor.  
In the presence of $dS_4$ isometry, only the $00$ and $55$ Einstein equations are independent, 
\begin{eqnarray}
H^2-n(y) n''(y)-n'(y)^2 = \frac{1}{4M_5^3}\frac{n(y)^2}{3}\left(\frac{1}{2}\Sigma'(y)^2+V(\Sigma)+\Lambda_5\right)\label{00},\\
n'(y)^2-H^2=\frac{1}{4M_5^3}\frac{n^2(y)}{6}\left(\frac{1}{2}\Sigma'(y)^2-V(\Sigma)-\Lambda_5\right)\label{55}.
\end{eqnarray}
Here and in the rest of the paper the $'$ will always denote a derivative with respect to the explicitly mentioned argument of the function. For example, $n'(y) $ and $n'(z)$ will denote $\frac{dn(y)}{dy}$ and  $\frac{dn(z)}{dz}$ respectively.
The Einstein equations above have to be supplemented with BC's \footnote{We are considering the extra dimension to be a $S^1/(Z_2\times Z_2^\prime)$ orbifold which gives rise to the BC's mentioned here.},
\begin{eqnarray}\label{jumpeq1}
\lim\limits_{\epsilon\rightarrow 0}\left[\frac{n'(y)}{n(y)}\right]_{-\epsilon}^{+\epsilon} = -\frac{V_0}{12 M_5^3},\\ \label{jumpeq2}
\lim\limits_{\epsilon\rightarrow 0}\left[\frac{n'(y)}{n(y)}\right]_{L-\epsilon}^{L+\epsilon} = -\frac{V_L}{12 M_5^3}.
\end{eqnarray}

\subsubsection{Gravitational Fluctuations}
The inflationary  $dS_4$ foliation necessarily ``warps'' the extra dimension, even when there is no bulk energy-momentum tensor. Thus the KK spectrum is also expected to be different from the non-inflationary 4D Lorentz-invariant case, with $\text{Mink}_4$ foliation. General gravitational fluctuations around the metric \eqref{bgmetric}, contains the graviton $h_{\mu\nu}(x,y)$ and, in presence of the second boundary, the radion $\Pi(x,y)$. We will show in Appendix \ref{appkk} that the linearized equation of motion for the spin-2 graviton and spin-0 radion decouple for a general warp factor $n(y)$. Thus postponing the discussion of radion to a later subsection,  we focus only on the 4D graviton and its KK modes for now. These fluctuations can be parametrized at the linearized level as,
\begin{equation}
ds^2=-n(y)^2dt^2+n(y)^2a(t)^2d\vec{x}^2+dy^2+h_{\mu\nu}(x,y)dx^\mu dx^\nu,
\end{equation}
where $\mu,\nu$ denote 4D indices $t, \vec{x}$.
In the above we have chosen $h_{5\mu}=0$ by a suitable gauge transformation, and $h_{\mu\nu}$ satisfies transverse and traceless conditions,
\begin{equation}
\nabla_\mu h^{\mu\nu}=0;\hspace{3em} h^\mu_{\hspace{0.5em}\mu}=0.
\end{equation}
The equation of motion for the graviton can be obtained as \cite{Garriga:1999bq,Langlois:2000ns,Karch:2000ct} (for a derivation see Appendix \ref{appkk})
\begin{equation}
\square_{dS} h_{\mu\nu}(x,y)+n^2(y)\partial_y^2h_{\mu\nu}(x,y)-2n'(y)^2h_{\mu\nu}(x,y)-2n(y)n''(y)h_{\mu\nu}(x,y)-2H^2h_{\mu\nu}(x,y)=0,
\end{equation} 
where $\square_{dS}=g^{\mu\nu}\nabla_\mu\nabla_\nu$ is the laplacian operator for $dS_4$ with $g_{\mu\nu}$ is the metric for $dS_4$ i.e. $ds_{4\text{D}}^2=g^{\mu\nu}dx_\mu dx_\nu = -dt^2+a^2(t)d\vec{x}^2$. To make the KK decomposition manifest we can redefine $h_{\mu\nu}=n^2\tilde{h}_{\mu\nu}$ to get
\begin{equation}
\square_{dS}\tilde{h}_{\mu\nu}(x,y)+n^2(y)\partial_y^2\tilde{h}_{\mu\nu}(x,y)+4n(y)n'(y)\partial_y\tilde{h}_{\mu\nu}(x,y)-2H^2\tilde{h}_{\mu\nu}(x,y)=0.
\end{equation}
 Expanding $\tilde{h}_{\mu\nu}$ into KK modes $\tilde{h}_{l,\mu\nu}(x)$ with profile $\chi_l(y)$
\begin{equation}\label{KKexpansion}
\tilde{h}_{\mu\nu}(x,y)=\sum_{l}\tilde{h}_{l,\mu\nu}(x)\chi_l(y),
\end{equation}
we get
\begin{eqnarray}
\square_{dS}\tilde{h}_{l,\mu\nu}(x)=(m^2+2H^2)\tilde{h}_{l,\mu\nu}(x)\\
n^2(y)\chi_l''(y)+4n(y)n'(y)\chi_l'(y)+m^2\chi_l(y)=0\label{profileeq}.
\end{eqnarray}
These describe $\tilde{h}_{l,\mu\nu}$ as spin-2 particles with mass $m$ in $dS_4$. The form taken by the bulk profile is most clear in the analog 1D ``quantum mechanics'' coordinate system, where eq. \eqref{profileeq} has the same form as Schroedinger equation with some potential determined by the warp factor $n(y)$ \cite{Randall:1999vf}. To achieve this, we can do a variable change $n(y)\frac{d}{dy}=\frac{d}{dz}$ and a field redefinition
\begin{equation}\label{fieldredef}
\chi_l(z) = n^{-\frac{3}{2}}(z)\psi_l(z)
\end{equation}
to get,
\begin{equation}\label{waveeq}
%-\frac{1}{2}\frac{d^2}{dz^2}\psi_l(z)+\left(\frac{3}{8}\left(\frac{n'(z)}{n(z)}\right)^2+\frac{3}{4}\frac{n''(z)}{n(z)}-\frac{V_0}{4}\delta(z)\right)\psi_l(z)=\frac{m^2}{2}\psi_l(z)
-\frac{1}{2}\frac{d^2}{dz^2}\psi_l(z)+\left(\frac{3}{8}\left(\frac{n'(z)}{n(z)}\right)^2+\frac{3}{4}\frac{n''(z)}{n(z)}%-\frac{V_0}{16M_5^3}\delta(z)
\right)\psi_l(z)=\frac{m^2}{2}\psi_l(z).
\end{equation}
We note that the effect of the bulk scalar $\Sigma$ comes only through the dependence on the warp factor $n(y)$ given via eqs. \eqref{00} and \eqref{55}. 
This is because the spin-0 fluctuations of $\Sigma$ cannot mix with spin-2 $h_{\mu\nu}$ at the linearized level. The zero mode profile for $m=0$ in eq. \eqref{waveeq} can be obtained for a general warp factor $n(z)$ with $\psi_0(z)\propto n^{\frac{3}{2}}(z)$.

\subsection{Semi-Infinite Extra Dimension}

We now specialize to the case  in which there is only one boundary at $y=0$, housing the inflaton. 
In this case, the radion is no longer in the spectrum. We therefore drop the stabilizing GW fields, and for simplicity consider vanishing 5D bulk cosmological constant. Then the warp factor $n(y)$ satisfying eqs. \eqref{00}, \eqref{55} and KK graviton profile obeying eq. \eqref{waveeq} simplifies significantly as we now demonstrate.
\subsubsection{Background Solution and the Horizon}
The solution to eq. \eqref{00} and eq. \eqref{55} along with BC eq. \eqref{jumpeq1} and normalization $n(y=0)=1$ is then given by
\begin{equation}\label{warpfactor} 
n=1 - H y,
\end{equation}
with $V_0=24 M_5^3 H > 0$. We see that the presence of the inflationary vacuum energy, characterized by $H\neq 0$, has ``warped'' the extra dimension giving rise to a horizon at $y=H^{-1}$ \cite{Kaloper:1998sw,Kaloper:1999sm}. 

\paragraph{Horizon Temperature.}
The temperature of the horizon can be found by studying the near horizon geometry. A variable change $\mathcal{Y}=H^{-1}-y$ shows that the line element transverse to the boundary becomes identical to a Rindler metric,
\begin{equation}
ds^2=-H^2\mathcal{Y}^2dt^2+d\mathcal{Y}^2.
\end{equation}
The temperature of this Rindler horizon can be found by the standard method of going to Euclidean time and demanding regularity of the metric at the horizon,
\begin{equation}
T_{\text{horizon}}=\frac{H}{2\pi}.
\end{equation}
This is same as the Gibbons-Hawking temperature of $dS_4$ space \cite{Gibbons:1977mu}. At first, this  coincidence of these temperatures is not clear. 

We gain insight by considering the case with negative 5D cosmological constant $\Lambda_5=-24M_5^3 k^2$ in the bulk, which corresponds to the RS2 set-up \cite{Randall:1999vf}, but with de-tuned boundary tension giving rise to $dS_4$ foliation rather than $\text{Mink}_4$ foliation.
The bulk equations \eqref{00} and \eqref{55} can again be solved \cite{Kaloper:1999sm,Kim:1999ja,Nihei:1999mt},
\begin{equation}\label{adswarpfactor}
n(y)=\cosh(ky)-\frac{\sqrt{H^2+k^2}}{k}\sinh(k y).
\end{equation}
With this warp factor, we again see the presence of a horizon with an identical near horizon geometry as before and horizon temperature $T_{\text{horizon}}=\frac{H}{2\pi}$. This can be interpreted as the temperature of the 4D CFT dual to RS2, as follows from the ``hot'' AdS/CFT correspondence. The CFT in this case is being heated by the $dS_4$ Gibbons-Hawking temperature due to 4D inflation. For aspects of such ``hot'' AdS/CFT correspondence, see \cite{Marolf:2010tg} and references therein.
 
\subsubsection{KK Graviton Wavefunction}

For now, let us return to the technically simpler case with vanishing $\Lambda_5$ (at the loss of holographic insight).
Using the explicit form of the warp factor eq. \eqref{warpfactor} in eq. \eqref{waveeq} we obtain
\begin{equation}\label{waveeq2}
\frac{d^2}{dz^2}\psi_l(z)+(m^2-\frac{9}{4}H^2+\frac{V_0}{8M_5^3}\delta(z))\psi_l(z)=0, 
\end{equation}
where the coordinate $z$ is defined by $e^{-zH}=1-H y$ \footnote{The delta function $\delta(z)$ originates because of the $R^1/Z_2$ quotient of an infinite extra dimension to obtain a semi-infinite extra dimension in the present case.}. Thus the horizon has been pushed to $z=\infty$, whereas the $y=0$ boundary resides at $z=0$. 

Remarkably, there is a $m=0$ normalizable and localized graviton mode,
\begin{equation}\label{zeromode}
\psi_0(z)\propto e^{-\frac{3}{2}H z},
\end{equation}
corresponding to a finite 4D effective Planck scale, $M_4$, as we will detail later.  This is similar to the RS2 graviton localization giving an effective 4D gravity despite the infinite extra dimension, but here the localization relies on 4D inflation, $H \neq 0$. Intuitively, the horizon provides a second boundary cutting off the infinite extra dimension.
We also see that for $m\neq 0$ there is a mass gap of $3H/2$ and a continuum of modes for $m>3H/2$. These modes are non-normalizable and their profile in the extra dimension is sinusoidal. This mass gap aligns nicely with the fact that a massive spin-2 particle in $dS_4$ has to obey the Higuchi bound \cite{Higuchi:1986py} $m^2\geq 2H^2$, which can be derived just by  unitarity of the 4D theory. (In inflationary scenarios where the $dS_4$ isometries are significantly broken, the Higuchi bound can be evaded and it is consistent to have spin-2 particles with $m^2 < 2H^2$ \cite{Bordin:2018pca}.)
It should be mentioned that for $\Lambda_5 < 0$, some of these features persist and have been pointed out in the literature, for e.g. \cite{Garriga:1999bq,Karch:2000ct,Langlois:2000ns}.

In this paper we will be interested in inflaton NG mediated by massive particles (with or without spin) having a \textit{discrete} spectrum. Hence to discretize the continuum modes with $m>\frac{3H}{2}$ above, we need to reintroduce a second boundary \textit{before} the horizon is reached at $y=H^{-1}$. We turn to this next.

\subsection{Introduction of the Second Boundary}

When the second boundary is introduced, the KK graviton wave function $\psi_l(z)$ has to obey two BC's so that the KK continuum above becomes discretized.
Furthermore the radion is a physical degree of freedom and we have to stabilize it. 

\subsubsection{Radion Mass and Stabilization}
We first set $\Lambda_5=0$ and ask what happens in the absence of a GW field. Using the metric solution, subject to $dS_4$ ansatz, given in eq. \eqref{warpfactor}, and the jump equations \eqref{jumpeq1} and \eqref{jumpeq2} the radius of the extra dimension is determined
\begin{equation}
L = \frac{V_L+V_0}{V_L H},
\end{equation}
even in the absence a GW field. Note that we need to have $V_L<0$ for there to be no horizon formed between the two boundaries. However, this $dS_4$-symmetric configuration is unstable as we discuss now.

We can parametrize the linearized radion fluctuation as \cite{Csaki:2000zn}
\begin{equation}
ds^2 = -n(y)^2(1-2\Pi(x,y))dt^2+n(y)^2a(t)^2(1-2\Pi(x,y))d\vec{x}^2+(1+2\Xi(x,y))dy^2.
\end{equation}
Although we have two seemingly independent functions, $\Pi(x,y)$ and $\Xi(x,y)$ to denote the radion, the perturbed $0i$ Einstein equations force $\Xi(x,y)=2\Pi(x,y)$ \cite{Csaki:2000zn}. Then the perturbed 55 Einstein equation, along with the background solution, gives the linearized radion equation of motion (for a derivation see Appendix \ref{appkk}),
\begin{equation}\label{radioneom1}
\square_{dS}\Pi+4 H^2 \Pi = 0.
\end{equation}
We see that the radion has a tachyonic mass $m_{r}^2=-4H^2$, signalling instability \cite{Frolov:2003yi}. 

For the case of $\Lambda_5<0$, it is still true that $m_r^2=-4H^2$ \cite{Gen:2000nu,Binetruy:2001tc,Chacko:2001em}. Although this can again be deduced by considering the perturbed Einstein equations, we can get the same result from a ``simple'' holographic insight. To this end we calculate the radius of the extra dimension via a similar procedure as above. Using eq. \eqref{adswarpfactor} and eqs. \eqref{jumpeq1}, \eqref{jumpeq2} we get,
\begin{equation}
\tanh(kL)=\frac{V_0+V_L}{24M_5^3k+\frac{V_0V_L}{24M_5^3k}}.
\end{equation} 
To have a solution to the above equation we need $V_L<-24M_5^3 k$. 
Now let us write down an effective potential for the canonically normalized radion  field $\Pi_c$ (which is proportional to $\Pi$) on this $dS_4$ symmetric background. 
As the Goldstone boson of spontaneous conformal symmetry breaking of the CFT dual to the bulk dynamics, the only possible conformally invariant form of the radion potential is,
\begin{equation}
V_r(\Pi_c)=H^2\Pi_c^2 +\lambda \Pi_c^4.
\end{equation}
A conformal coupling of the radion to the 4D Ricci scalar, $\mathcal{L}\supset -\frac{1}{12}R \Pi_c^2+\cdots$ fixes the mass term above with $R=12 H^2$ for $dS_4$. A quartic coupling $\lambda$ is also expected to be present whenever the tension on the second boundary $V_L$ is not equal to the tuned RS1 value of $-24M_5^3k$, with $\text{sgn}(\lambda)$ being fixed by $\text{sgn}(V_L+24M_5^3k)$ (see e.g. \cite{Chacko:2013dra}). Since we needed $V_L<-24M_5^3 k$ in the present case, we have $\lambda<0$. If we now expand around the correct minima of $\Pi_c$, we get back the identical tachyonic mass $m_r^2=-4H^2$ as before.

The tachyonic radion necessitates the presence of some stabilization mechanism. %For that purpose, in the following let us reintroduce the GW field $S$. 
Since the tachyonic instability is $\sim -\mathcal{O}(H^2)$ and we are interested in having $m_{KK}\sim H$ for observability of NG, the stabilization will necessarily have an $\mathcal{O}(1)$ backreaction on the geometry. However, this makes the analysis technically more difficult since we have to solve coupled field equations for the GW field and the metric. Fortunately, as we discuss below, this analysis simplifies in a near-horizon approximation and  yields important qualitative insights.

\subsubsection{Near-horizon Analysis of Stabilization}

We begin by noting that the observability of KK gravitons of the compactified (2-boundary) scenario is tightly constrained by purely 4D considerations: Boltzmann suppression $\sim e^{-\pi \mu}$ for large $m$ and the Higuchi lower bound $m>\sqrt{2}H$ (following from unitarity). The former can be seen by an explicit calculation of the bispectrum due to 
KK graviton exchange which we carry out in Appendix \ref{appA} and detail further in Section \ref{kkgravng}. In Fig. \ref{fig:spin2} we plot the function $f_2(\mu)$ (defined in eq. \eqref{f2def}) which characterizes the strength of NG due to KK graviton exchange and from there it is evident that significant Boltzmann suppression kicks in soon as $m$ gets bigger than $\frac{3H}{2}$ \footnote{The apparent divergence of $|f_2(\mu)|$ as $\mu\rightarrow 0$ is actually absent in the full bispectrum, since in the limit of $\mu\rightarrow 0$ only the real part of $f_2(\mu)$ contributes which remains finite.}. 
Hence to have an observable NG mediated by KK gravitons, we need to have their masses within a narrow window about $\frac{3H}{2}$. 

Fortunately, we saw above that in the absence of a second boundary, there is a horizon at a finite proper distance $y=H^{-1}$ and a continuum of KK graviton modes starting precisely at $\frac{3H}{2}$. In the presence of a second boundary this continuum spectrum must turn into a discrete one. However if the warp factor $n(y)$ on the second boundary is $\ll1$, i.e. if the second boundary is placed just in front of a ``would-be'' horizon, we expect to get back a finely discretized spectrum of KK gravitons starting around $m = 3H/2$, thereby avoiding significant Boltzmann suppression. To show this we first write the linearized warp factor near the second boundary as,
\begin{equation}
n(\mathcal{\varepsilon}) \approx H\varepsilon,
\end{equation}
where $\varepsilon$ is the coordinate along the extra dimension and the horizon is reached as $\varepsilon\rightarrow 0$ \footnote{$\varepsilon$ can be related to $y$ once we know the warpfactor along the entire extra dimension, but in this paper we will be solving for the warp factor only near the second boundary.}. Again with the coordinate transformation $-n\frac{d}{d\varepsilon}=\frac{d}{dz}$ we can write the equation of motion for the wave function in the analog 1D ``quantum mechanics'' coordinate system, as in eq. \eqref{waveeq}, 
\begin{equation}\label{waveeq3}
-\psi_l''(z)+(\frac{9H^2}{4}-m^2)\psi_l(z)=0.
\end{equation}
This form is identical to what we had in the absence of the second boundary, eq. \eqref{waveeq2}, without the delta function source \footnote{Note that by a slight abuse of notation we used the same variable $z$ in both eqs. \eqref{waveeq2} and \eqref{waveeq3} whereas they match only very near the horizon}. However, unlike that case, now the $z$ coordinate does not extend to $\infty$, but rather to some finite, but large (in units of $1/H$) value. Since $n(\varepsilon) \approx e^{-zH}$, by making the warp factor on the second boundary smaller, we can make the size of the ``box'' bigger in the analog quantum mechanics problem, and thereby decreasing the spacing between the KK modes. 

Having motivated the need for a near-horizon boundary, we now have to ask whether such a configuration can actually be stabilized. To this end, we reintroduce a GW field $\Sigma$ with a bulk mass $m_\Sigma$. Then
its extra dimensional profile follows the bulk equation of motion with $dS_4$ ansatz,
\begin{equation}\label{GW}
\Sigma''(y)+4n'(y)/n(y)\Sigma'(y)=m_\Sigma^2\Sigma(y).
\end{equation}
We have to solve the coupled set of equations \eqref{00}, \eqref{55} and \eqref{GW} to obtain a consistent background solution, which is difficult to do in general. 
But in the near-horizon limit we are interested in we can solve the coupled set of equations perturbatively in $H\varepsilon$. We take the ansatz for the warp factor and the profile to be
\begin{eqnarray}\label{metansatz}
n(\varepsilon) = 
a_1 H\varepsilon + a_2 H^2\varepsilon^2/2 + a_3 H^3\varepsilon^3/3 +\cdots\\ %+ a_4 \delta^4/
%4 + a_5 \delta^5/5\\\label{GWansatz}
\Sigma(\varepsilon) = b_0 + b_1 H\varepsilon + b_2 H^2\varepsilon^2/2 + b_3 H^3\varepsilon^3/3 +\cdots. %+ b_4 \delta^4/4 + b_5 \delta^5/5
\end{eqnarray}
We will focus on the regime $H^{-1}\gg \varepsilon >\varepsilon_c$ with $\varepsilon_c$ being the location of the second boundary. 
The solution to eqs. \eqref{00},\eqref{55} and \eqref{GW} is given by,
\begin{eqnarray}\label{nearhormetric}
n(\varepsilon)=H\varepsilon -\frac{1}{72}v^2m_\Sigma^2 H\varepsilon^3+\cdots \\\label{nearhorGW}
\Sigma(\varepsilon)=\sqrt{4M_5^3}\left(v+\frac{1}{10}v m_\Sigma^2\varepsilon^2+\cdots\right),
\end{eqnarray}
where $v$ is some constant fixed by the BC's on the GW field. Note when the stabilizer is absent i.e. $v=0$, we get back the near horizon behavior given in \eqref{warpfactor} with $\varepsilon = H^{-1}-y$. 

Now let us analyze the radion equation of motion. For this we have to consider the fluctuation $\sigma(x,y)$ of the background GW field $\Sigma(y)$, since the former can mix with the radion. We can go through the perturbed Einstein equations once again to get the equation of motion for the radion \eqref{radioneom},
\begin{equation}\label{radioneom2}
\frac{1}{n^2}\square_{dS} \Pi=-\Pi''-2n'\Pi'/n+4((\frac{n'}{n})^2-\frac{n''}{n})\Pi+2\frac{\Sigma''}{\Sigma'}(\Pi'+2n'\Pi/n)-6H^2\Pi/n^2.\\
%e^{2 A}\square_{dS} \Pi=-4H^2 e^{2A}\Pi+\frac{1}{4M_5^3}\left(\frac{2}{3}S'^2F+\frac{1}{3}\sigma S''-\frac{1}{3}\sigma'S'\right).
\end{equation}
In the above, we have used the bulk equation,
\begin{equation}\label{mixing}
\Pi'+2\frac{n'}{n}\Pi=\frac{1}{12M_5^3}\sigma \Sigma',
\end{equation}
to eliminate $\sigma$ dependence in eq. \eqref{radioneom2}. We have also used $'$ to denote $\frac{\partial}{\partial y}$. We can now plug in the background solutions given by eqs. \eqref{nearhormetric} and \eqref{nearhorGW} in eq. \eqref{radioneom2} to find, 
\begin{equation}\label{radionsol}
\Pi(x,\varepsilon)\propto (H\varepsilon)^{\frac{1}{2}\pm \nu}+\cdots
\end{equation}
with $\nu=\sqrt{\frac{9}{4}-\frac{m_r^2}{H^2}}$. In the absence of the GW field we had to have $m_r^2=-4H^2$ to satisfy the radion equation of motion \eqref{radioneom1}. Now, with  the stabilizer such a constraint has disappeared since eq. \eqref{radionsol} is a near-horizon solution for arbitrary $m_r$, and with a suitable choice of BC's we can make $m_r^2>0$. Thus by studying the near horizon geometry, we have shown how to stabilize the radion in presence of a GW field. 

We see that we can stabilize the second boundary arbitrarily close to the would-be horizon, and that this results in a finely-spaced spin-2 KK spectrum beginning arbitrarily close to 
$m = 3H/2$. This demonstrates that the KK modes need not suffer large Boltzmann suppressions in their NG contributions. 

 Before proceeding further, let us make a comment about the KK spectrum during and after inflation which we denote by $M_{KK}^{\text{inf}}$ and $M_{KK}^{\text{today}}$ respectively. We should note that the observed values of the SM gauge couplings suggest, within the orbifold GUT paradigm, an extra dimension with size $M_{KK}^{\text{today}}\sim 10^{14}$ GeV \textit{today}. On the other hand, an inflationary Hubble scale $H_{\text{inf}}\sim 5\times 10^{13}$ GeV is allowed by data and motivated by high-scale inflation models. Thus we see that it is entirely possible to have $M_{KK}^{\text{today}}\sim H_{\text{inf}}$. But this alone does not guarantee that we will see interesting and observable NG signals due to KK states, since for that we actually need $M_{KK}^{\text{inf}}\sim H_{\text{inf}}$. Here is where the stabilizing GW scalar plays a crucial role by determining the size of the extra dimension, in the low curvature regime given by 
	\begin{equation}
 M_{KK}^{\text{today}}\underset{\mathrm{H_{\text{today}}\approx 0}}{\sim} m_\Sigma \ln(v_1/v_2),
	\end{equation}
where the $v_{1,2}$ are the VEVs of the GW scalar on the two boundaries. Since we are considering  $M_{KK}^\text{today}\sim 10^{14}$ GeV \textit{today}, this implies 
$m_\Sigma\sim 10^{14} \text{GeV}\sim H_{\text{inf}}$. With $m_\Sigma\sim H_{\text{inf}}$, our near-horizon analysis then shows that there is no obstruction in stabilizing 
the non-inflaton boundary near the would-be horizon, guaranteeing $M_{KK}^\text{inf}\approx 3H_{\text{inf}}/2$. We can contrast this with what would have happened if either 
$M_{KK}^{\text{today}} \gg H_{\text{inf}}$ or $M_{KK}^{\text{today}} \ll H_{\text{inf}}$. In the former case, we would have 
$m_{\Sigma} \gg H_{\text{inf}}$, and Hubble expansion 
would generically\footnote{We can see this in eqs. \eqref{nearhormetric} and \eqref{nearhorGW}, where the near-horizon expansion for large $m_{\Sigma} \gg H$ clearly requires parametrically small $\epsilon_c$, which in turn requires $\epsilon_c$-level tuning of parameters to stabilize. Generically there is no such tuning and hence no near-horizon stabilization for $m_{\Sigma} \gg H$.}
be subdominant in the stabilization dynamics from the time of inflation all the way until today, so that $M_{KK}^{\text{inf}}  \approx  M_{KK}^{\text{today}}  \gg H_{\text{inf}}$, and seeing the GUT 
states would be highly Boltzmann suppressed. In the latter case, we would have $m_{\Sigma} \ll H_{\text{inf}}$, and the $\sigma$ mixing terms in eq. \eqref{mixing} would be negligible, so 
we would approximately have the unstabilized result that the radion would be tachyonic if the boundary is near the horizon. 
Thus, the rough coincidence $M_{KK}^{\text{today}}  \sim H_{\text{inf}}$ plays a critical role in allowing us to see the GUT states in NG.

\subsection{Inflationary Couplings}

\subsubsection{Wavefunction of KK Graviton on Inflationary Boundary}
To determine the coupling of the KK graviton to the inflaton, localized at the $y=0$ boundary, we will need the wavefunction of the KK graviton at $y=0$. However, we have argued above that to stabilize the radion, the backreaction of the GW field on the metric will typically be $\mathcal{O}(1)$, so that this will also affect the KK graviton wavefunction at the $\mathcal{O}(1)$ level.
In order to precisely calculate this we would have to extend our near-horizon analysis of the last subsection to the entire extra-dimensional interval. It would be interesting to find some analytic means of doing this (non-perturbative in $H$) , but as yet we do not have such an analysis. The superpotential approach taken in \cite{DeWolfe:1999cp} may be useful in this regard. Here, we will simply estimate the KK graviton wavefunctions by ignoring the backreaction completely, but assign an $\mathcal{O}(1)$ uncertainty to this estimate.

We therefore proceed by beginning with the metric for the single boundary set-up, eq. \eqref{warpfactor}, but taking the extra dimension to simply be cut off by the location of the second boundary at say $y_c$ before reaching the horizon.  This neglects the backreaction of the requisite stabilization of the second boundary, as discussed above. In terms of the coordinate $z$ defined by $n\frac{d}{dy}=\frac{d}{dz}$, we have $n(z)=e^{-zH}$ with the extra dimension ranging from $z=0$ to $z_c=-\frac{1}{H}\ln(1-Hy_c)$. Furthermore in this coordinate system the profile of KK modes, obeying eq. \eqref{waveeq2}, is sinusoidal. The orthonormality condition is given by
\begin{equation}\label{orthonormality}
2M_5^3\int_0^{z_c} dz\psi_l^*(z)\psi_m(z)=\frac{M_4^2}{2}\delta_{lm},
\end{equation}
where the numerical factor is chosen to ensure that the
4D action is given by $\frac{M_4^2}{2}\int d^4 x R_4$ with $R_4$ and $M_4$ being the 4D Ricci scalar and the 4D Planck scale respectively. As will be explained below $M_4$ differs from the standard Planck scale $M_{\text{pl}}=2.4\times 10^{18}$ GeV by some $\mathcal{O}(1)$ amount due to inflationary dynamics. However in the end, this difference will not be important for us because the final strength of KK graviton NG \eqref{f2def} will be dependent on $M_4$ only via the tensor-to-scalar ratio $r$. Thus only the observational upper bound on $r$ \cite{Akrami:2018odb}, rather than an actual knowledge of $M_4$, will be important.
%zero mode graviton has canonical normalization in 4D. 
Thus in the $z$ coordinate system, the wavefunction behaves as if it is in a flat extra dimension and after normalization, it will carry the usual ``$\frac{1}{\sqrt{\text{Volume}}}$ dilution factor'' (see e.g. \cite{Sundrum:2005jf}). On the boundary containing the inflaton, the wavefunction is given by,
\begin{equation}\label{wavefunc}
\psi_l(z=0)\sim \frac{1}{\sqrt{H z_c}}\sim \frac{1}{\sqrt{-\ln(n(y_c))}}
\end{equation}
with $z_c$ being the ``volume'' of the extra dimension. In the above, we have used the relation $z_c=-\frac{1}{H}\ln(1-Hy_c)=-\frac{1}{H}\ln(n(y_c))$. As we show in the following, the strength of the coupling  between the inflaton and the KK graviton is proportional to $\psi_l(z=0)$ and eq. \eqref{wavefunc} shows that such a coupling is only logarithmically suppressed when we place the second boundary very near a would-be horizon. Crucially, this will allow us to get an observable NG signal mediated by KK gravitons, without paying a large wavefunction suppression in the coupling strength. This \textit{logarithmic} suppression, however, seems unavoidable in our set up. This is because, although decreasing the size of the extra dimension will increase the overlap between the inflaton and the KK graviton---hence increasing the coupling---it
is also expected to make the KK gravitons heavier and thereby we will incur \textit{exponential} Boltzmann suppression in NG signals. Let us now write down the explicit inflaton-KK graviton coupling using which we estimate the strength of NG mediated by KK gravitons in the subsequent discussion.

\subsubsection{Coupling of KK Graviton to the Inflaton}
At the linear order, the graviton fluctuations couple to the energy-momentum tensor of the inflaton in the standard way, namely,
\begin{equation}\label{infKKcoupling}
S_{\text{int}}=\int d^4x \frac{\delta S_{inf}}{\delta g_{\mu\nu}}h_{\mu\nu} = -\frac{1}{2}\int d^4x \sqrt{-g}T^{\mu\nu}_{inf}h_{\mu\nu}
\end{equation}
where we have used the definition of the energy momentum tensor $T^{\mu\nu}_{inf}=-\frac{2}{\sqrt{-g}}\frac{\delta S_{inf}}{\delta g_{\mu\nu}}$. Since we are using the convention that the warp factor $n(y=0)=1$ on the inflationary boundary, using the expansion \eqref{KKexpansion} and eq. \eqref{fieldredef} we can simplify eq. \eqref{infKKcoupling} as,
\begin{equation}
-\frac{1}{2}\sum_{l=0}^{\infty}\int d^4x \sqrt{-g} T^{\mu\nu}_{inf}\tilde{h}_{l,\mu\nu}\psi_l(0).
\end{equation}
Upon canonically normalizing \footnote{We will continue to denote the canonically normalized KK gravitons by the same variable $\tilde{h}_{l,\mu\nu}$ to simplify the notation.} the massless 4D graviton and the KK modes, from the above we get,
\begin{equation}
-\frac{1}{M_4}\int d^4x \sqrt{-g} T^{\mu\nu}_{inf}(\tilde{h}_{0,\mu\nu}+\tilde{h}_{1,\mu\nu}\psi_1(0)+\cdots)
\end{equation}
We have set $\psi_0(0)=1$ without loss of generality and focused only on the first (i.e. the lightest) KK mode for concreteness. Finally using the fact that we are in the gauge $h^\mu_{\hspace{0.5em}\mu}=0$ we get the coupling between the inflaton and the KK graviton,
\begin{equation}\label{inf-kk}
-\frac{1}{M_4}\int d^4x \sqrt{-g}\partial^\mu\phi\partial^\nu\phi(\tilde{h}_{1,\mu\nu}\psi_1(0)+\cdots).
\end{equation}

\subsubsection{Estimate of NG Mediated by KK Graviton}

Let us now give a quick estimate of the NG mediated by KK graviton using the coupling in eq. \eqref{inf-kk}. We can expand the inflaton in terms of the background $\phi_0$ and the fluctuation $\xi$ to get, 
\begin{equation}
-\frac{\psi_1(0)}{M_4}\tilde{h}^{\mu\nu}_1\partial_\mu\phi\partial_\nu\phi = -\frac{\psi_1(0)}{M_4}\tilde{h}^{\mu\nu}_1\left(\partial_\mu\phi_0\partial_\nu\phi_0+2\partial_\mu\phi_0\partial_\nu\xi+\partial_\mu\xi\partial_\nu\xi\right)
\end{equation}
The first term gives a small tadpole, which can be shifted via a field redefinition without affecting the relevant couplings significantly, whereas the second term after using $\nabla_\mu h^{\mu\nu}=0$ gives,
\begin{equation}
-\frac{2 \psi_1(0)}{M_4}H\dot{\phi}_0 \xi \tilde{h}^{00}_1.
\end{equation}
Hence the relevant couplings are given by,
\begin{equation}\label{kkcoupling}
-\frac{2 \psi_1(0)}{M_4}H\dot{\phi}_0 \xi \tilde{h}_1^{00} -\frac{\psi_1(0)}{M_4}\tilde{h}_1^{\mu\nu} \partial_\mu \xi \partial_\nu \xi.
\end{equation}
From the above we can get a quick estimate of the parametric strength of NG defined in eq. \eqref{F}, 
\begin{equation}\label{kkngestimate}
F\sim \frac{\psi_1(0)}{M_4}\times \frac{\psi_1(0)\dot{\phi}_0}{M_4}\times \frac{\dot{\phi}_0}{H^2} \sim \frac{\dot{\phi}_0^2}{M_4^2H^2} \times \psi_1(0)^2,
\end{equation}
while a detailed form containing the momentum dependence as in eq. \eqref{fnl_qsfi2} will be given in Sec. \ref{kkgravng}. 

The quantity $\frac{\dot{\phi}_0^2}{M_4^2H^2}$ differs by an ${\cal O}(1)$ factor from its standard value of $2\epsilon$ in a purely 4D set-up \cite{Giudice:2002vh} where $\epsilon\equiv -\frac{\dot{H}}{H^2}$. To understand why, note that ordinarily after compactification, the 4D EFT is generally an expansion in $~ E/m_{KK} < 1$. In cosmology a characteristic energy scale is  $E \sim H$, and in the present context we seek $m_{KK}$ comparable to $H$ for KK visibility in NG. Therefore we should expect ${\cal O}(1)$ corrections relative to the leading 4D predictions. 
To see this explicitly we can consider the inflaton EOM with the potential $V_0$ (which we previously approximated as a constant) on the inflationary boundary,
\begin{equation}
\ddot{\phi}_0+3H\dot{\phi}_0+\frac{dV_0}{d\phi}=0.
\end{equation}
The Friedman equation, following from eq.  \eqref{jumpeq1} (with the warp factor $n(y)=1-Hy$) reads as,
\begin{equation}
\frac{1}{2}\dot{\phi}_0^2+V_0=24M_5^3H.
\end{equation}
Using the relation \eqref{M4} between $M_5, M_4$ and $H$, $M_4^2=4M_5^3L\left(1-HL+\frac{H^2L^2}{3}\right)$,  and using the usual definition of  $\epsilon=-\frac{\dot{H}}{H^2}$, 
one sees that $\frac{\dot{\phi}_0^2}{M_4^2H^2}\neq 2\epsilon$.

It will be useful to write the quantity $\frac{\dot{\phi}_0^2}{M_4^2H^2}$ in terms of the tensor to scalar ratio, $r$, in order to estimate the strength of the KK graviton mediated NG signals in Sec. \ref{kkgravng}. In our set-up, the scalar power spectrum will be unaffected, to the leading order in slow-roll parameters, by the presence of the extra dimension since the inflaton fluctuations are localized on the boundary \cite{Maartens:1999hf,Giudice:2002vh}. Hence $r$ is given by,
\begin{equation}\label{r}
r \equiv \frac{P_{T,k}}{P_{S,k}} = 8\frac{ \dot{\phi}_0^2}{H^2M_4^2}.
\end{equation}
In the above, we have used the tensor power spectrum,
$P_{T,k} = \frac{H^2}{M_4^2}\frac{4}{k^3}$,
and the scalar power spectrum $P_{S,k}=\frac{H^4}{\dot{\phi}_0^2}\frac{1}{2k^3}$.

Now we come back to the estimate of $F$ in eq. \eqref{kkngestimate}. As argued earlier, the wavefunction suppression above is quite mild and hence the KK graviton mediated NG is expected to be of the order of $f_{\text{NL}}\sim r < 10^{-1}$. While inaccessible by future large-scale structure surveys \cite{Alvarez:2014vva}, such a level of NG should be potentially observable by 21-cm experiments probing the dark ages \cite{Meerburg:2016zdz} if we have a high scale inflation scenario with $H\lesssim 10^{13}$ GeV. We conclude this section by checking whether such a large value of $H$ is consistent within our set-up.

\paragraph{Cutoff of 5D Gravity.} 
To have quantum gravity corrections under control, we should have $V_0<M_5^4$. To check that, first we recall the graviton zero mode profile given in eq. \eqref{zeromode}, 
\begin{equation}
\psi_0=e^{-\frac{3H}{2} z},
\end{equation}
and use the normalization condition in eq. \eqref{orthonormality} to get,
\begin{equation}\label{M4}
M_4^2=4M_5^3L\left(1-HL+\frac{H^2L^2}{3}\right)
\end{equation}
In the above we have assumed the warp factor is given by $n(y)=1-Hy$ ignoring the backreaction of the stabilizer field. Taking $L\approx 1/H$ we get,
\begin{equation}
\frac{V_0}{M_5^4} = 24(4/3)^{1/3}\times\left(H/M_4\right)^{2/3} 
%= 24(8/3)^{1/3}\times \left(r/8\right)^{1/3}\left(H^2/\dot{\phi}_0\right)^{2/3}
\ll 1.
\end{equation}

\section{Gauge Theory States}\label{kkgauge}

Whereas observing a KK graviton resonance via NG would be striking, it would be even more so if we see the accompanying signatures of massive gauge bosons. The latter can arise naturally in our set up as the KK modes of the bulk unified gauge fields. The observability of NG mediated by such KK gauge bosons will depend both on their masses and coupling to the inflaton. Interestingly, we will see below that the set-up with a near-horizon second boundary, chosen above to give us
 $m_{\text{KK}}^{\text{graviton}}\sim\mathcal{O}(H)$, also  yields $m_{\text{KK}}^{\text{gauge}}\sim\mathcal{O}(H)$. Thus in such a set-up, the cosmological production of KK gauge bosons will not be significantly Boltzmann suppressed and the observability of KK gauge boson mediated NG will depend solely on their coupling strength to the inflaton. We start by analysing the mass spectrum of the KK gauge bosons.

\subsection{KK Analysis of 5D Gauge Theory}
Let us focus on the case of a bulk $U(1)$ gauge theory which is sufficient for finding free-field profiles of the self-interacting bulk non-Abelian gauge theory.  
The 5D action is given by,
\begin{equation}
S_{U(1)} = \int \sqrt{-G}G^{MN}G^{PQ}F_{MP}F_{NQ},
\end{equation}
where 
\begin{equation}
ds^2 = G_{MN}dx^M dx^N  = n(y)^2 g_{\mu\nu}dx^\mu dx^\nu + dy^2.
\end{equation}
$G_{MN}$ corresponds to the 5D metric governing the line element \eqref{bgmetric} while $g_{\mu\nu}$ denotes the metric of $dS_4$ in flat Poincare coordinates. $M,N$ and $\mu,\nu$ run over the 5D and 4D indices respectively. 

By a suitable gauge transformation and orbifold BC's, $A_y$ can be eliminated from the physical spectrum. The equation of motion for the gauge boson is then given by,
\begin{equation}\label{eomgauge}
\nabla^\nu F_{\nu\mu}(x,y)+\partial_y(n^2\partial_y A_\mu(x,y)) = 0,
\end{equation}
where $\nabla$ denotes the covariant derivative w.r.t $dS_4$. Via a KK decomposition,
\begin{equation}
A_\mu(x,y) = \sum_{l}A_{l,\mu}(x)\vartheta_l(y),
\end{equation}
the equation of motion \eqref{eomgauge} can be rewritten as,
\begin{eqnarray}\label{gaugeeom4d}
\nabla^\nu F_{l,\nu\mu}(x) = m_l^2A_{l,\mu}(x), \\ \label{gaugeprofile1}
\partial_y(n^2\partial_y \vartheta_l(y))+m_l^2 \vartheta_l(y) = 0.
\end{eqnarray}
Eq. \eqref{gaugeeom4d} describes the usual 4D equation of motion for a massive/massless gauge field in $dS_4$, whereas eq. \eqref{gaugeprofile1} governs the profile of the KK gauge boson in the extra dimension. With our earlier variable change, $n(y)\frac{d}{dy} = \frac{d}{dz}$, and a field redefinition 
\begin{equation}
\vartheta_l(y) = n^{-1/2}(y)\tilde{\vartheta}_l(y),
\end{equation}
 we can rewrite the eq. \eqref{gaugeprofile1} as,
\begin{equation}\label{gaugeprofile2}
\tilde{\vartheta}_l''(z)+\left(\frac{1}{4}\left(\frac{n'(z)}{n(z)}\right)^2-\frac{1}{2}\frac{n''(z)}{n(z)}+m_l^2\right)\tilde{\vartheta}_l(z) = 0.
\end{equation}
The zero mode profile can be obtained for a general warp factor $n(z)$ and is given by $\tilde{\vartheta}_0(z)\propto n^{1/2}(z)$.
\paragraph{Mass Spectrum.}To analyze the KK gauge boson mass spectrum we can proceed in a manner similar to the case of the KK graviton. For a moment let us go to the case where the second boundary is absent, so that the extra dimension ends in the horizon $z = \infty$. Then the warp factor \eqref{warpfactor} is given by $n(z) = e^{-zH}$ and correspondingly eq. \eqref{gaugeprofile2} reduces to,
\begin{equation}\label{gaugeprofile3}
\tilde{\vartheta}_l''(z)+\left(m_l^2-\frac{H^2}{4}\right)\tilde{\vartheta}_l(z) = 0.
\end{equation}
First, note that for $m_l=0$ we will have a zero mode whose profile is given by,
\begin{equation}\label{zeromodegauge}
\tilde{\vartheta}_0(z)\propto e^{-\frac{Hz}{2}}.
\end{equation}
Furthermore, we will have a continuum of KK gauge bosons above $m_l>\frac{H}{2}$. This particular lower bound is significant because if we now place the second boundary very near, but before we reach the horizon, the KK modes will get discretized and the lightest of the KK modes will have masses $\approx \frac{H}{2}$. These lightest KK modes can mediate observable NG without significant Boltzmann suppression.

\paragraph{Wavefunction of KK Gauge Boson on Inflationary Boundary.}
The coupling of the KK gauge boson to the inflaton, localized at the $y=0$ boundary, is determined by the wavefunction of the KK gauge boson at $y=0$. To find the wavefunction, in principle, we have to solve eq. \eqref{gaugeprofile2} after the backreaction of the stabilizing GW field has been taken into account. However, 
using the same reasoning as in the previous section, we will simply estimate the KK gauge boson wavefunction by ignoring the effects of backreaction completely and assigning an $\mathcal{O}(1)$ uncertainty in our estimate. Under this approximation the KK gauge boson profile, obeying eq. \eqref{gaugeprofile3} between the two boundaries at $z=0$ and $z=z_c=-\frac{1}{H}\ln(n(y_c))$, behaves as if it is in a flat extra dimension. Hence the profiles will be sinusoidal and when normalized they will carry the usual ``$\frac{1}{\sqrt{\text{volume}}}$ dilution factor''. Thus on the boundary containing the inflaton, the KK gauge boson wavefunction is given by,
\begin{equation}
\tilde{\vartheta}'_l(z=0)\sim \frac{1}{\sqrt{H z_c}}\sim \frac{1}{\sqrt{-\ln(n(y_c))}}.
\end{equation}
As for KK gravitons, the fact that this wavefunction suppression is only logarithmic, will allow us to get an observable NG.

\subsection{Contribution of KK Gauge Boson to NG}
\paragraph{Cutoff of 5D Gauge Theory.}
To explain the observed smallness of the slow roll parameter $\eta\sim 10^{-2}$, in the following, we will impose an (approximate) shift symmetry on the inflaton. This implies that the inflaton-gauge boson couplings will necessarily involve higher dimension operators suppressed by some field theory cutoff scale $\Lambda_\text{inf}$. For consistency of the derivative expansion in $\frac{(\partial\phi)^2}{\Lambda_\text{inf}^4}$, we require $\Lambda_\text{inf}>\sqrt{\dot{\phi}_0} \sim 60 H$ \cite{Creminelli:2003iq}. Furthermore the 5D gauge theory, being non-renormalizable, will be valid only below a certain energy scale $\Lambda_\text{gauge}$. A naive dimensional analysis shows that such a scale is given by,
\begin{equation}
\Lambda_{\text{gauge}} \sim \frac{1}{N}\frac{16\pi^2}{g_5^2},
\end{equation}
where $N$ is the number of colors if the gauge group is of $SU$ type. 
Note the gauge zero mode profile \eqref{zeromodegauge} is flat in the $y$ coordinate system defined in eq. \eqref{bgmetric}. Hence the 5D gauge coupling $g_5$ will be related to the 4D gauge coupling $g_4$ via eq. \eqref{unification} (using $L\sim H^{-1}$),
\begin{equation}
\frac{1}{H g_5^2}\sim \frac{1}{g_4^2}.
\end{equation}
In the above, we have used the fact that in the near-horizon set-up we are working in, the size of the extra dimension is $\sim \frac{1}{H}$. Taking $m_{\text{KK}}\sim H$ we get,
\begin{equation}
\frac{m_{\text{KK}}}{\Lambda_{\text{gauge}}}\sim \frac{g_4^2 N}{16\pi^2}.
\end{equation}
As an example, with $N\sim 5$ and the gauge coupling at the unification scale, $\frac{g_4^2}{4\pi}\sim \frac{1}{40}$, (see Fig. \ref{fig:thresold-correction-orbifold} ) we get,
\begin{equation}
\Lambda_{\text{gauge}} \sim 100 H.
\end{equation} 
Since $\Lambda_{\text{gauge}} \gtrsim   \sqrt{\dot{\phi}_0} \sim 60 H$ we can simply take $\Lambda_\text{inf}\sim\Lambda_{\text{gauge}}\sim 100 H$ to have the derivative expansion in $\frac{(\partial\phi)^2}{\Lambda_\text{inf}^4}$ under control. From now on we will use $\Lambda$ to denote this common cut-off scale. Alternatively, we can switch to the effective theory of inflation \cite{Cheung:2007st} in which the scale $\dot{\phi}_0$ does not appear, in which case a lower $\Lambda$, and consequently larger NG, is allowed. We will not pursue this direction further in this paper. 

For the choice $H\sim 5\times 10^{13}$ GeV, motivated by the observed (approximate) unification of the gauge couplings (Fig. \ref{fig:thresold-correction-orbifold}), we have $  V_{\text{inf}}^{1/4} \sim 10^{16} \text{GeV} \gtrsim \Lambda_{\text{gauge}} \sim 5\times 10^{15}\text{GeV}$.  This suggests that the
5D gauge theory may need to be UV completed a little below the inflationary vacuum energy scale. 
This does not conflict with obtaining effective inflaton-gauge interactions suppressed only by $\Lambda$ if these are mediated by massive states, as explored in Ref. \cite{Kumar:2017ecc}.

The interaction between the inflaton and KK gauge boson is constrained by the fact that we take the inflaton to be a singlet under the bulk gauge group. As a consequence, if we restrict ourselves to tree level ``in-in'' diagrams  (for the sake of observability), the KK gauge boson must also be singlet under the broken gauge group to mediate a non-zero NG. To illustrate this restriction, we now discuss two well motivated scenarios where the unified gauge groups in the bulk are respectively $SO(10)$ and $SU(5)$.

\subsubsection{$SO(10)$ GUT in the Bulk}
In this case, with Neumann inflationary-BC's for the the SM subgroup gauge fields and Dirichlet inflationary-BC's for the remaining $SO(10)/$SM gauge fields, as well as Neumann 
 BC's on all $SO(10)$ gauge bosons on the near-horizon boundary (preserving the entire $SO(10)$ symmetry there), we end up with only SM gauge field zero-modes after KK reduction. We need only respect the preserved SM gauge invariance in coupling on the inflationary boundary. Under the SM gauge symmetry one
  of the broken generators is a  singlet, corresponding to $B-L$ symmetry. The associated gauge field, which we simply denote by $A_{\mu}$, can therefore be coupled to the SM-singlet inflaton, unconstrained except for spacetime symmetries. While $A_{\mu}$ has no zero-mode, its KK excitations can thereby mediate NG.

\paragraph{Inflationary Couplings of the B-L Gauge Boson.}
Our choice of Dirichlet BC on the inflationary boundary and the absence of restrictions imposed by gauge invariance give the following lowest dimension operators that give the \textit{leading} contributions to NG, 
\begin{multline}
\mathcal{L}_{\text{inf-gauge}}\supset \frac{c_1}{\Lambda^3}( \partial_yA_\nu) (\partial_y A_\mu) \nabla^\mu\nabla^\nu \phi + 
\frac{c_2}{\Lambda^4}(\nabla\phi)^2 (\partial_yA_\mu)^2 +
\frac{c_3}{\Lambda^4}(\nabla_\mu\phi \partial_y A^\mu)^2 +\\
+ \frac{c_4}{\Lambda^4}(\nabla\phi)^2\nabla_\mu\phi \partial_yA^\mu 
+\frac{c_5}{\Lambda^4} \partial_yA^{\mu} \nabla_\mu\phi \partial_y A_\nu \partial_y A^\nu+\cdots
\end{multline}
In the above $c_i$'s are some coefficients of $\mathcal{O}(1)$. We have omitted a term of the type $\rho_1\nabla_\mu \phi \partial_y A_\mu$, since its effects are negligible for $\rho_1 \lesssim 1$, which is natural. 

To obtain the couplings required for estimating the bispectrum, we expand the inflaton field, $\phi = \phi_0(t) +\xi(t,\vec{x})$ as before. It can be seen that $\mathcal{L}_{\text{inf-gauge}}$ contains a gauge boson tadpole coming from the term with coefficient $c_4$. Such a tadpole can be removed by a field redefinition, without significantly affecting the relevant couplings for the parameter choice we will be focusing on. $\mathcal{L}_{\text{inf-gauge}}$ also contains several terms of the form $A_0^2$ and $A_\mu^2$. Such mass corrections also will not give a large effect within the same parameter choice. Keeping up to cubic order in fluctuations, $\mathcal{L}_{\text{inf-gauge}}$ is then given by
\begin{multline}\label{infgauge}
\frac{c_1}{\Lambda^3}\left(A'^\mu A'^\nu(\partial_\mu\partial_\nu\xi-\Gamma^\alpha_{\mu\nu}\partial_\alpha\xi)\right) + \frac{2c_2}{\Lambda^4}\dot{\phi}_0\dot{\xi}A'^2_\mu+\frac{2c_3}{\Lambda^4}\dot{\phi}_0A'^0\partial_\mu\xi A'^\mu \\
+\frac{2c_4}{\Lambda^4}\dot{\phi}_0\dot{\xi}\partial_\mu\xi A'^\mu +\frac{c_4}{\Lambda^4}\dot{\phi}_0 A'^0(\partial_\mu\xi)^2+\frac{2c_4}{\Lambda^4}\dot{\phi}_0^2\dot{\xi}A'^0 \\
+\frac{c_5}{\Lambda^4}\dot{\phi_0}A'^0 A'^2_\nu.
\end{multline}
In the above the $'\equiv\frac{\partial}{\partial y}$.
\paragraph{Estimates of NG.}
Eq. \eqref{infgauge} contains interactions that can give rise to single, double and triple exchange diagrams for NG based on the number of gauge boson propagators, see Fig. \ref{fig:diagrams}. Let us estimate each of these in turn, 
\begin{align}\label{gaugesingle}
F^\text{single}&\sim c_4^2\times\frac{\dot{\phi}_0^4}{\Lambda^8}\times \vartheta_1'(0)^2,\\ \label{gaugedouble}
F^\text{double}&\sim \left(c_2\text{ or }c_3\right)\times c_4^2\times\frac{\dot{\phi}_0^4}{\Lambda^8}\times \frac{\dot{\phi}_0^2}{\Lambda^4}\times \vartheta_1'(0)^4,\\ \label{gaugetriple}
F^\text{triple}&\sim c_5\times c_4^3\times \frac{\dot{\phi}_0^6}{\Lambda^{12}}\times \frac{\dot{\phi}_0^2}{\Lambda^4}\times \vartheta_1'(0)^6.
\end{align}
In the above we have kept the $\mathcal{O}(1)$ coefficients $c_i$'s to be explicit about the particular couplings contributing to each of the diagrams. The fact that $F^\text{single}$ is sensitive to a very high power of the cut-off scale $\Lambda$, namely $\sim \Lambda^{-8}$ implies that NG will be significantly suppressed (and, possibly unobservable) if $\Lambda\gg \sqrt{\dot{\phi}_0}$. However, we saw above that the 5D gauge theory breaks down at a scale $\Lambda_{\text{gauge}}\gtrsim\sqrt{\dot{\phi}_0}$, hence taking $\Lambda\sim \Lambda_{\text{gauge}}$ we can have $F^\text{single}\lesssim 1$. For the same scenario, $F^\text{double}$ and $F^\text{triple}$ are somewhat smaller than $F^\text{single}$ because of the extra suppressions due to $\vartheta_1'(0)\lesssim 1$ and $\frac{\dot{\phi}_0^2}{\Lambda^4}\lesssim 1$, but they can still be observable for favorable values of $\Lambda$ and $\vartheta_1'(0)$. In Section \ref{kkgaugeng} we will give the detailed form of NG mediated by the single exchange diagram using the results from our previous work \cite{Kumar:2017ecc}.

\subsubsection{$SU(5)$ GUT in the Bulk}
In this case, with Neumann inflationary-BC's for the the SM subgroup gauge fields and Dirichlet inflationary-BC's for the $X,Y$ gauge fields, as well as Neumann 
near-horizon BC's on all $SU(5)$ gauge bosons (preserving the entire $SU(5)$ symmetry there), we again end up with only SM gauge field zero-modes after KK reduction.

Two scenarios can arise now: (a) the SM gauge group remains unbroken at energies $\sim H$,
and (b) through the presence of a non-minimal Higgs-curvature coupling ${\cal L} \supset c R_4\mathcal{H}^\dagger \mathcal{H}, ~ c>0$ the electroweak symmetry gets spontaneously broken at inflationary scales $\sim H$.  After inflation ends, the curvature effect of such a non-minimal coupling decreases rapidly and electroweak symmetry gets restored until the SM temperature falls below $\sim 100$ GeV. This is the scenario of ``heavy-lifting'' \cite{Kumar:2017ecc}.
For case (a) there are massive gauge singlets (under the unbroken SM gauge group), namely the KK excitations of hypercharge gauge boson, $B_{l,\mu}$. However because $U(1)_Y$ is \textit{un}broken, the quadratic mixing between the inflaton and $B_{l,\mu}$---necessary for a non-zero bispectrum---will be highly suppressed. Hence the resulting bispectrum is expected to be unobservably small. But this does not mean that a bulk $SU(5)$ GUT will not have any NG signature, since for case (b), there will be a massive $Z$ boson. This will have $\mathcal{O}(H)$ mass for $\mathcal{O}(1)$ non-minimal coupling (i.e. $c\sim1$) and can couple to the inflaton with appreciable strength to mediate observable NG. This type of scenario has been discussed at length in \cite{Kumar:2017ecc} and hence, we will not pursue it here further.

\section{Detailed Form of NG Mediated by Spin-2}\label{kkgravng}
\begin{figure}[h]
	\centering
	\includegraphics[width=0.6\linewidth]{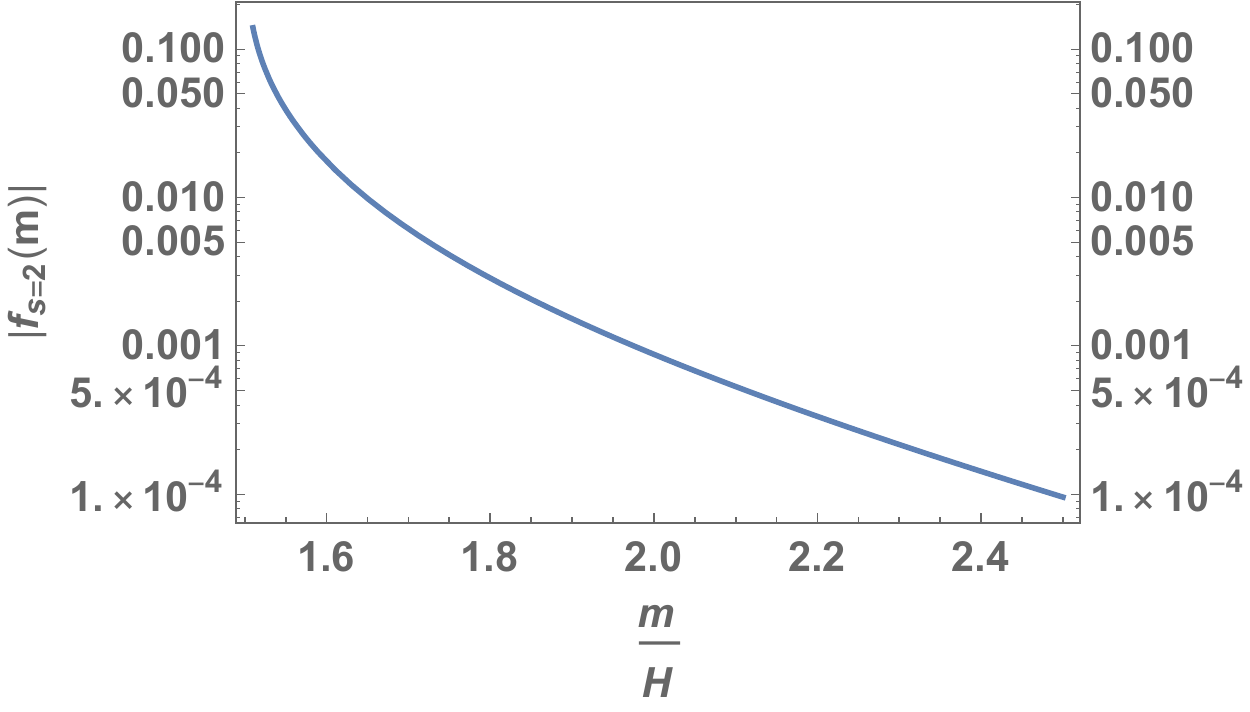}
	\caption{Strength of NG mediated by spin-2 KK graviton for tensor-to-scalar ratio $r=0.1$ and KK wavefunction on inflationary boundary $\psi(0)=1$. Such strengths for the range of masses shown are observable within cosmic variance (see Section \ref{ininreview})}
	\label{fig:spin2}
\end{figure}

In the following we focus on the single exchange diagram, given in Fig. \ref{fig:diagrams}, for KK graviton mediated NG. Since the inflaton-KK graviton couplings are $\sim M_4$ suppressed, the double and triple exchange diagrams will be more suppressed compared to the single exchange diagram. The couplings relevant for computing this diagram can be obtained from eq. \eqref{kkcoupling} \footnote{We will drop the subscript in $\psi_1(0)$ for brevity.}, 
\begin{equation}
-\frac{2 \psi(0)}{M_4}H\dot{\phi}_0 \xi h^{00} -\frac{\psi(0)}{M_4}h^{\mu\nu} \partial_\mu \xi \partial_\nu \xi,
\end{equation}
and the resulting NG is given by eq. \eqref{f2}, (using eq. \eqref{r} to write $\frac{\dot{\phi}_0^2}{M_4^2H^2}=\frac{r}{8}$)
\begin{multline}
\frac{5}{18}F_{\text{KK Graviton}}^{\text{single}}=\frac{5}{18}\psi(0)^2\frac{r}{8}\times (\cos^2\theta-\frac{1}{3})\frac{\sqrt{\pi}}{8 (1+4\mu_2^2)^2\cosh(\pi\mu_2)}\times  \\
\left(A(\mu_2)(1+i\sinh\pi\mu_2)
\left(\frac{k_3}{k_1}\right)^{3/2+i \mu_2} + (\mu_2\rightarrow-\mu_2)\right)\\\label{f2def}
\equiv \left(\cos^2\theta-\frac{1}{3}\right) \times \left( f_2(\mu_2)\left(\frac{k_3}{k_1}\right)^{\frac{3}{2}+i\mu_2}+f_2(-\mu_2)\left(\frac{k_3}{k_1}\right)^{\frac{3}{2}-i\mu_2}\right),
\end{multline}
with 
\begin{equation}
A(\mu)=(-27 + 120 i \mu + 152 \mu^2 - 32 i \mu^3 + 16 \mu^4) \Gamma(5/2 + i \mu) \Gamma(-i \mu) 2^{-2 i \mu},
\end{equation}
and $\mu_2=\sqrt{\frac{m^2}{H^2}-\frac{9}{4}}$. The factor of $\frac{5}{18}$ is present to conform with the definition of $f_{\text{NL}}$ parameter in eq. \eqref{fnl}. We plot $|f_2(\mu_2)|$ in Fig. \ref{fig:spin2} to illustrate the strength of NG signal mediated by KK gravitons. 
Using the discussion following eq. \eqref{zeromode}, we see that  as the non-inflaton boundary approaches the would-be horizon the effective mass parameter $\mu_2\rightarrow 0$ \footnote{This feature persists even when there is a bulk cosmological constant, see e.g. \cite{Garriga:1999bq,Langlois:2000ns,Karch:2000ct}.}. We will encounter an identical feature for the case of gauge bosons in the following.

\section{Detailed Form of NG Mediated by Spin-1}\label{kkgaugeng}
\begin{figure}[h]
	\centering
	\includegraphics[width=0.6\linewidth]{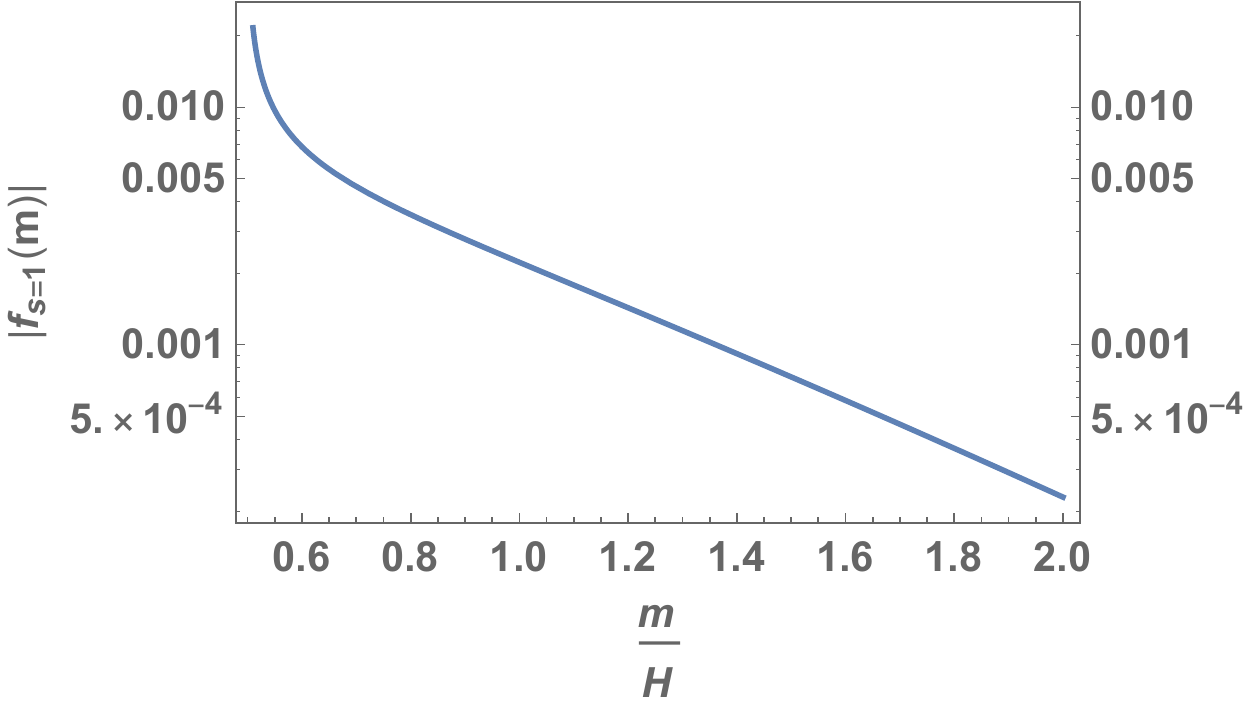}
	\caption{Strength of NG mediated by spin-1 KK gauge bosons for inflaton-KK mixing $\rho=0.3$ and derivative of KK wavefunction on the inflationary boundary $\vartheta'(0)=1$. Such strengths for the range of masses shown are observable within cosmic variance (see Section \ref{ininreview})}
	\label{fig:spin1}
\end{figure}

In the following we focus on the single exchange diagram for KK gauge boson mediated NG. The double and triple exchange diagrams are expected to be somewhat suppressed compared to the single exchange diagram, as we estimated in eqs. \eqref{gaugesingle}-\eqref{gaugetriple}. For the single exchange diagram, the relevant couplings that give an angular dependence that is characteristic of a spin-1 exchange, can be obtained from  eq. \eqref{infgauge},
\begin{equation}
+\frac{\rho}{\dot{\phi}_0}\dot{\xi}\partial_i\xi A'^i +\rho\dot{\xi}A'^0,
\end{equation} 
where $\rho=\frac{2c_4}{\Lambda^4}\dot{\phi}_0^2$ gives the inflaton-KK gauge boson mixing. The resulting strength of NG has been calculated in \cite{Kumar:2017ecc} and is given by,
	\begin{multline}\label{singlezeft}
\frac{5}{18}F^{\text{single}}_{\text{KK Gauge Boson}}=\frac{5}{18}\left(\frac{\rho}{m}\right)^2\frac{1}{16\pi}\sin^2\theta\Gamma(\frac{3}{2}+i\mu_1)\Gamma(\frac{3}{2}-i\mu_1)\cosh(\pi\mu_1)\vartheta'(0)^2 \times\\
\left((7-5i\mu_1+16\mu_1^2+4i\mu_1^3)\Gamma(\frac{3}{2}+i\mu_1)^2\Gamma(-2-2i\mu_1)(1-i\sinh(\pi\mu_1))\left(\frac{k_3}{k_1}\right)^{\frac{5}{2}+i\mu_1}+(\mu_1\rightarrow-\mu_1)\right)\\
\equiv
\sin^2\theta \times \left( f_1(\mu_1)\left(\frac{k_3}{k_1}\right)^{\frac{5}{2}+i\mu_1}+f_1(-\mu_1)\left(\frac{k_3}{k_1}\right)^{\frac{5}{2}-i\mu_1}\right),
\end{multline}
where $\mu_1=\sqrt{\frac{m^2}{H^2}-\frac{1}{4}}$ and $\vartheta'(0)$ is the derivative of the wavefunction of the KK gauge boson on the inflationary boundary. As in the case of KK gravitons,  we plot $|f_1(\mu_1)|$ in Fig. \ref{fig:spin1} to illustrate the strength of NG signal mediated by KK gauge bosons. 
Using the discussion following eq. \eqref{zeromodegauge}, we see that as the non-inflaton boundary approaches the would-be horizon the effective mass parameter $\mu_1\rightarrow 0$, similarly to the case of KK gravitons above. Furthermore, it can be seen using eq. \eqref{gaugeprofile2} (which is valid for a general warp factor $n(z)$), that the above feature persists even if there is a negative bulk cosmological constant. In fact, for such a case of $\approx AdS_5$ geometry in the bulk, the non-inflaton boundary being very close to the horizon is holographically dual to a strongly-interacting and confining matter sector,  which due to the inflationary Gibbons-Hawking temperature is heated to be close to its confinement-deconfinement phase transition. We expect that there is some deep (holographic) significance to $\mu_{1,2} \rightarrow 0$ at this transition, but we have not found it beyond just direct 5D computation. A simpler and deeper understanding would also allow us to predict if $\mu \rightarrow 0$ applies to more general spins in more general models. We hope to come back to this issue in future work.

\section{Conclusion and Future Directions}\label{conclusion}
The observation that the SM gauge couplings become approximately equal to each other at $M_U\sim 10^{14}$ GeV hints at the exciting possibility of grand unification around that scale. Although such a scale is too high to directly probe using terrestrial colliders, an inflationary era in the primordial universe offers a unique opportunity in that regard. Since the inflationary Hubble scale $H$ can be as big as $M_U$, inflationary spacetime can produce such $M_U$-scale GUT states \textit{on-shell} which can then decay into inflatons and give distinct, non-analytic NG contributions to the spectrum of primordial curvature perturbations, from which one can extract the masses and spins of such GUT states. 

Motivated by their simplicity and the ease of suppressing proton decay, we have focused on orbifold GUTs and studied the strength of such NG signals mediated by KK GUT gauge bosons  and KK gravitons.
An optimal scenario is identified where the extra dimension is stabilized, via a Goldberger Wise scalar, close to the onset of a bulk event horizon such that there is a discrete KK spectrum but with small enough splittings that their production does not suffer significant Boltzmann suppression. In such a scenario, we have found that \textit{both} the KK gravitons and KK gauge bosons can mediate potentially observable NG  allowing for a unique and direct probe of orbifold GUTs. A (near) future discovery of primordial gravity waves from inflation---implying $H\sim M_U$---combined with a discovery of both spin-1 and spin-2 mediated NG signals, and an absence of higher spin signals (hinting at the absence of composite or stringy effects during inflation) would make a strong observational case for an orbifold GUT structure during inflation.

There remain various interesting directions for future work. From Figs. \ref{fig:spin2} and \ref{fig:spin1} we see that the strengths of NG---characterized by the $f_{\text{NL}}$ parameter---mediated by KK gravitons and KK gauge bosons are typically $f_{\text{NL}}<0.1$. Although such a level of NG can be potentially observable using futuristic 21-cm cosmology experiments, they will be difficult to detect via upcoming Large Scale-Structure surveys which will mostly probe $f_{\text{NL}}\sim \mathcal{O}(1)$ (See \cite{Alvarez:2014vva} for a summary). Hence it is important to look for variations in our set-up in which stronger NG can be obtained. Let us briefly mention two separate possibilities in which one can get potentially larger NG mediated by KK gravitons and KK gauge bosons respectively.

We saw in Sec. \ref{kkgrav} that the inflationary couplings of the KK gravitons are model independent and suppressed by the 4D Planck scale, $M_4$. Hence to get a larger KK graviton mediated NG, we have to increase the strength of this gravitational coupling. Fortunately, Randall Sundrum models \cite{Randall:1999ee,Randall:1999vf} already provide an example where the 4D Planck scale gets warped down, in the presence of a bulk 5D cosmological constant, as one moves towards the infrared (IR) boundary. Thus with the inflaton localized on the IR boundary or in the bulk one can expect to have stronger coupling between the inflaton and the KK graviton. However, one has to be careful as to whether the large inflationary vacuum energy stored on the IR boundary can backreact significantly on the geometry and take into account the effect of that on the KK graviton mode functions. 

Interactions between the KK gauge bosons and the inflaton involve higher dimension operators suppressed by the cutoff scale $\Lambda$. This is due to the shift symmetry of the inflaton and the Dirchlet boundary conditions on the non-SM gauge fields on the inflationary boundary in Fig. \ref{fig:extra-dim-geometry}. Since we described the inflationary dynamics in the paradigm of single-field slow-roll inflation we had to impose the constraint $\Lambda>\sqrt{\dot{\phi}}_0\sim 60H$. However it is possible that the single-field slow-roll paradigm is not an appropriate description of inflationary dynamics and in particular some unknown new physics comes in at energies $\Lambda_{\text{EFT}}\ll \sqrt{\dot{\phi}}_0$. To capture the effects of such new physics, we can write an effective field theory (EFT), valid $\lesssim H$, for the inflaton which is a Goldstone of the time translation breaking \cite{Cheung:2007st}. Within such an EFT one can parametrize the inflaton interactions systematically in an expansion in $\frac{H}{\Lambda_{\text{EFT}}}$. With $\Lambda_{\text{EFT}}\ll \sqrt{\dot{\phi}}_0$ one can then obtain larger KK gauge boson mediated NG signals. 

We have seen that a complete description of a stabilization mechanism of the extradimensional set-up with two boundaries involves solving the coupled Einstein equations for the stabilizer field and the metric simultaneously. In general, this is difficult to do analytically. In this paper, we have done a near-horizon analysis of stabilization by which we can systematically solve the coupled equations perturbatively, and the warp factor $n(y)$ very near the second boundary is determined that way. This allows us to compute NG but with $\mathcal{O}(1)$ uncertainties. It would therefore be very useful to find an analytic way of solving the coupled equations in the entire extra dimension. The superpotential approach taken in \cite{DeWolfe:1999cp} can help in this regard. In that case, we could calculate the precise inflationary couplings of the KK modes by determining their profile in the entire extra dimension and thereby obtain a more precise calculation of the NG they mediate. 
\newpage
\section*{Acknowledgements}
The authors would like to thank Nima Arkani-Hamed, Junwu Huang, Mark Van Raamsdonk for helpful discussions. This research was supported in part by the NSF under Grant No. PHY-1620074 and by the Maryland Center for Fundamental Physics (MCFP).

\appendix 
\section{KK Reduction of the Graviton-Radion System}\label{appkk}
The linearized gravitational fluctuations to the background metric \eqref{bgmetric} can be characterized by,
\begin{multline}
ds^2 = -n(y)^2(1-2\Pi(x,y))dt^2+n(y)^2a(t)^2(1-2\Pi(x,y))d\vec{x}^2+(1+4\Pi(x,y))dy^2\\+h_{\mu\nu}(x,y)dx^\mu dx^\nu,
\end{multline}
where $h_{\mu\nu}$ and $\Pi(x,y)$ denote the graviton and the radion fluctuations respectivey. We have chosen a gauge such that $\nabla_\mu h^{\mu\nu}=0=h^\mu_{\hspace{0.5em}\mu}$. In the following we derive the linearized equation of motion for the graviton and the radion from the perturbed Einstein equations,
\begin{equation}
\delta R_{MN}=\frac{1}{4M_5^3}\delta\tilde{T}_{MN},
\end{equation}
where $\tilde{T}_{MN}=T_{MN}-\frac{1}{3}g_{MN}T^A_{\hspace{0.5em}A}$ with $T_{MN}$ being the bulk stress-energy tensor. Our approach will be similar to \cite{Csaki:2000zn} and we generalize their results appropriately to the case of a $dS_4$ foliation with $H\neq 0$.

For a generic metric fluctuation $\delta G_{MN}$, we can get the linearized perturbed Ricci tensor \cite{Higuchi:1986py},
\begin{multline}\label{perturbedricci}
\delta R_{MN}=\frac{1}{2}\left(\nabla_A\nabla_M\delta G^A_{\hspace{0.5em}N}+\nabla_A\nabla_N\delta G^A_{\hspace{0.5em}M}\right)-\frac{1}{2}\nabla_A\nabla^A\delta G_{MN}
-\frac{1}{2}\left(\nabla_N\nabla_M \delta G^A_{\hspace{0.5em}A}\right).
\end{multline}
To show that the graviton and the radion equation of motion decouple at the linearized level, we split $\delta R_{MN}$ into,
\begin{equation}
\delta R_{MN}=\delta R_{MN}^h + \delta R_{MN}^\Pi,
\end{equation}
where $\delta R_{MN}^{h(\Pi)}$ is linear in $h_{\mu\nu}(\Pi)$. Then using eq. \eqref{perturbedricci} and the identity,
\begin{equation}
[\nabla_A,\nabla_M] \delta G^A_{\hspace{0.5em}N}=\bar{R}_{BM}\delta G^B_{\hspace{0.5em}N}-\bar{R}^B_{\hspace{0.5em}NAM}\delta G^A_{\hspace{0.5em}B},
\end{equation}
we can derive,
\begin{eqnarray}
\delta R^h_{\mu 5}=0;\hspace{3em}\delta R^h_{5 5}=0.
\end{eqnarray}
This implies the $55$ and $5\mu$ Einstein equations can only contribute to the radion eq. of motion which we now derive. To do this first we evaluate,
\begin{eqnarray}
\delta R^F_{\mu 5}=3\partial_\mu \Pi'+6\frac{n'}{n}\partial_\mu \Pi,\nonumber\\
\delta R^F_{5 5} = -\frac{2}{n^2}\square_{dS}\Pi+4\Pi''+16\frac{n'}{n}\Pi'\nonumber,
\end{eqnarray}
where $\square_{dS}$ d'Alembertian for $dS_4$. In the above and the rest of this Appendix, $'\equiv \frac{\partial}{\partial y}$. We will also need the perturbed stress energy tensors,
\begin{eqnarray}
\delta\tilde{T}_{5\mu}=\partial_\mu \sigma \Sigma',\nonumber\\
\delta \tilde{T}_{55} = 2\Sigma'\sigma'+\frac{2}{3}\frac{dV(\Sigma)}{d\Sigma}\sigma+\frac{8}{3}V(\Sigma)\Pi\nonumber.
\end{eqnarray}
The GW field is expanded as $\Sigma(y)+\sigma(x,y)$ where 
$\sigma$ is the fluctuation of the background GW field $\Sigma$.
Then the $55$ Einstein equation gives the radion eq. of motion,
\begin{align}\label{radioneom}
\frac{1}{n^2}\square_{dS} \Pi=-\Pi''-2n'\Pi'/n+4((\frac{n'}{n})^2-\frac{n''}{n})\Pi+2\frac{\Sigma''}{\Sigma'}(\Pi'+2n'\Pi/n)-6H^2\Pi/n^2,
\end{align}
while the $5\mu$ Einstein equations give (after doing an integration to get rid of $\partial_\mu$),
\begin{equation}\label{radionmixing}
3\Pi'+6\frac{n'}{n}\Pi = \frac{1}{4M_5^3}\sigma \Sigma'.
\end{equation}
We can consider the special case of an unstabilized extra dimension where the GW field is absent. In that case eq. \eqref{radionmixing} simplifies to give $3\Pi'+6\frac{n'}{n}\Pi = 0$, so that \eqref{radioneom} becomes,
\begin{equation}
\frac{1}{n^2}\square_{dS} \Pi = 2((\frac{n'}{n})^2-\frac{n''}{n})\Pi-6H^2\Pi/n^2 \hspace{2em}(\text{No Stabilization}).
\end{equation}
Then using the background eqs. \eqref{00} and \eqref{55} we get,
\begin{equation}\label{tachradion}
\square_{dS} \Pi = -4H^2\Pi\hspace{2em}(\text{No Stabilization}),
\end{equation}
which shows that the radion gets a tachyonic mass of $-4H^2$ in absence of a stabilizing GW field.

Now let us study the $\mu\nu$ equations. We expect these to give the graviton eq. of motion, but first we have to show that the radion decouples from these equations. This can be done using the expressions,
\begin{eqnarray}\label{radionmunu}
\delta R_{\mu\nu}^\Pi = g_{\mu\nu}\square_{dS}\Pi-g_{\mu\nu}n^2(-24\Pi(n'/n)^2-6\Pi(n'/n)'-10\Pi'n'/n-\Pi''),\\\label{mixingeq.}
\delta \tilde{T}_{\mu\nu}^\Pi = -\frac{4}{3}V\Pi n^2g_{\mu\nu}+\frac{2}{3}\frac{dV}{d\Sigma}\sigma n^2 g_{\mu\nu},
\end{eqnarray}
where $g_{\mu\nu}$ is the metric for background $dS_4$ spacetime (without the $n(y)^2$ warp factor). Using the eqs. \eqref{radionmunu}, \eqref{mixingeq.} and \eqref{radioneom} we can derive
that $\delta R_{\mu\nu}^\Pi=\frac{1}{4M_5^3}\delta \tilde{T}_{\mu\nu}^\Pi$. Hence the $\mu\nu$ eqs. imply $\delta R_{\mu\nu}^h=\frac{1}{4M_5^3}\delta \tilde{T}_{\mu\nu}^h$, from which we will get the graviton eq. of motion. Thus we have decoupled the graviton-radion system. $\delta R_{\mu\nu}^h$ can be evaluated to be,
\begin{equation}
\delta R_{\mu\nu}^h = -\frac{1}{2n^2}\square_{dS}h_{\mu\nu}-\frac{1}{2}h_{\mu\nu}''-2(n'/n)^2h_{\mu\nu}+4\frac{H^2}{n^2}h_{\mu\nu}.
\end{equation}
Using $\delta\tilde{T}_{\mu\nu}^h=\frac{2V}{3}h_{\mu\nu}$ we finally arrive at the graviton eq. of motion,
\begin{equation}
\frac{1}{n^2}\square_{dS}h_{\mu\nu} +h_{\mu\nu}''-2(n'/n)^2h_{\mu\nu}-2n''/n h_{\mu\nu}-2H^2/n^2 h_{\mu\nu}=0.
\end{equation}

\section{NG Mediated by KK Graviton}\label{appA}
To calculate KK graviton mediated NG, we will need the mode functions of a massive spin-2 field in $dS_4$ \cite{Lee:2016vti} which we now derive. 
\subsection{Mode Functions for Helicity-0 Component of a Massive Spin-2 Field in $dS_4$ }

\paragraph{Helicity Decomposition.}The NG contribution that we are interested in involves quadratic mixing between the inflaton and the KK graviton. Since the inflaton is a scalar, only the scalar degree of freedom (DOF), or the helicity 0 component of a massive spin-2 particle in 4D, can be relevant. This DOF will come from metric fluctuation $h_{\eta\eta}$ and helicity 0 components of $h_{i\eta}$ and $h_{ij}$. To isolate the helicity 0 component from the 3-vector $h_{i\eta}$ we can write it as a gradient of a scalar and a divergenceless vector, in momentum space,
\begin{equation}
h_{i\eta}(\eta,\vec{k}) = \hat{k}_i h_V(\eta,\vec{k}) + \cdots,\\
\end{equation}
where we have omitted the divergenceless vector for brevity. To isolate the same from $h_{ij}$ we first note that to implement $h^\mu_{\hspace{0.5em}\mu}=0$ we can write, $h_{ij}=h^{\text{traceless}}_{ij}+\frac{1}{3}h_{\eta\eta}\delta_{ij}$, and then write the traceless part as,
\begin{equation}
h^{\text{traceless}}_{ij}(\eta,\vec{k})=\epsilon_{ij}(\vec{k})h_\text{T}(\eta,\vec{k})+\cdots,
\end{equation}
where $\epsilon_{ij}(\vec{k})=\frac{3}{2}(\hat{k}_i\hat{k}_j-\frac{1}{3}\delta_{ij})$ and $\cdots$ contain the helicity $\pm 1$ and $\pm 2$ fluctuations which we have not kept for brevity. In the above $\hat{k}_i$'s are unit vectors.

\paragraph{Mode Functions.}
We now focus on deriving the mode functions for $h_{\eta\eta}$ and $h_{T}$ which will be required for computing KK graviton mediated NG that will have a characteristic spin-2 angular dependence. First, from the eq. of motion $\square_{dS}h_{\eta\eta}=(m^2+2H^2)h_{\eta\eta}$ we get,
\begin{equation}
\partial_\eta^2h_{\eta\eta}+\frac{2}{\eta}\partial_\eta h_{\eta\eta}-\frac{4}{\eta}\partial_ih_{i\eta}-\frac{2}{\eta^2}h_{ii}+\frac{m^2/H^2-6}{\eta^2}h_{\eta\eta}-\partial_i^2h_{\eta\eta}=0.
\end{equation}
To convert the above into an eq. of motion involving only $h_{\eta\eta}$, we 
apply the constraints $h^\mu_{\hspace{0.5em}\mu}=0$ and
\begin{equation}
\nabla^\mu h_{\mu\eta}=\partial_\eta h_{\eta\eta}-\frac{1}{\eta}h_{\eta\eta}-\partial_ih_{i\eta}-\frac{1}{\eta}h_{ii}=0,
\end{equation}
to get,
\begin{equation}\label{hetaeta}
\partial_\eta^2h_{\eta\eta}-\frac{2}{\eta}\partial_\eta h_{\eta\eta}+\frac{m^2}{H^2\eta^2}h_{\eta\eta}-\partial_i^2h_{\eta\eta}=0.
\end{equation}
Using the constraint, 
\begin{equation}
\nabla^\mu h_{\mu i} = \partial_\eta h_{\eta i}-\frac{2}{\eta}h_{i\eta}-\partial_j h_{ij} = 0,
\end{equation}
we can obtain an \textit{algebraic} equation for $h_{ij}$,
\begin{equation}\label{hT}
\partial_\eta^2h_{\eta\eta}-\frac{4}{\eta}\partial_\eta h_{\eta\eta}+\frac{6}{\eta^2}h_{\eta\eta} = \partial_i\partial_j h_{ij}.
\end{equation}
Note the above equation is sufficient to determine the helicity-0 component of $h_{ij}$, i.e. $h_T$. Hence to summarize, by solving eqs. \eqref{hetaeta} and \eqref{hT} we will get the desired mode functions. To canonically quantize the spin-2 field we can follow the standard procedure as in the case of scalars. We write the fields $h_{\eta\eta}$ and $h_T$ in terms of linear combinations of the creation and destruction operators,
\begin{eqnarray}
h_{\eta\eta}(\eta,\vec{k}) = h_{k,0}(\eta)a_{\vec{k}}^\dagger +\bar{h}_{k,0}(\eta)a_{-\vec{k}},\\
h_{T}(\eta,\vec{k}) = h_{k,T}(\eta)a_{\vec{k}}^\dagger +\bar{h}_{k,T}(\eta)a_{-\vec{k}},
\end{eqnarray}
where $h_{k,0}(\eta),\bar{h}_{k,0}(\eta)$ and $h_{k,T}(\eta),\bar{h}_{k,T}(\eta)$ are solutions of eqs. \eqref{hetaeta} and \eqref{hT} respectively. In particular,
\begin{equation}
\bar{h}_{k,0}(\eta) = e^{i\pi/4}e^{-\pi \mu/2} \mathcal{N}_k (-k\eta)^{\frac{3}{2}}H_{i\mu}^{(1)}(-k\eta),
\end{equation}
and,
\begin{multline}
\bar{h}_{k,T}(\eta) = \frac{1}{12}e^{i\pi/4}e^{-\pi \mu/2}\mathcal{N}_k(-k\eta)^{-\frac{1}{2}}\nonumber\\\times\left(-6(2-i\mu)k\eta H_{i\mu-1}^{(1)}(-k\eta)+6(2+i\mu)k\eta H_{i\mu+1}^{(1)}(-k\eta)-(9-8k^2\eta^2)H_{i\mu}^{(1)}(-k\eta)\right)\label{transverse},
\end{multline}
where $\mathcal{N}_k=\sqrt{\frac{\pi}{6}}\frac{\sqrt{k}}{H}\frac{H}{m\sqrt{m^2/H^2-2}}$ is a normalization factor which can be derived by demanding the orthonormality of the mode functions \cite{Lee:2016vti}, and $\mu=\sqrt{m^2/H^2-9/4}$. 
\subsection{Calculation of the Single Exchange Diagram}
In this subsection we will be interested in computing the NG mediated by a single KK graviton exchange as in Fig. \ref{fig:diagrams} using the master formula \eqref{ininmasterformula} for computing an in-in expectation values. Our discussion here will be very brief and for a more detailed explanation of the set-up and the notation, we refer the reader to our previous work \cite{Kumar:2017ecc}. We will also momentarily work in $H=1$ units and restore $H$ in the final expression for NG in eq. \eqref{f2}.

The lagrangian relevant for the single exchange diagram can be obtained from eq. \eqref{kkcoupling},
\begin{equation}
\mathcal{L}=-\frac{2\psi_1(0)}{M_4}\eta^2\dot{\phi}_0\xi  h_{\eta\eta} - \frac{\psi_1(0)}{M_4}\eta^4 \partial_i\xi \partial_j\xi  \epsilon_{ij}h_T+\cdots.
\end{equation}
In the cubic term above we have kept only the spatial metric fluctuation $h_{ij}$, since that gives an angular dependence that is characteristic of a spin-2 exchange, and used its helicity-0 piece. The three point function corresponding to this single exchange diagram will consist of 4 diagrams, $I_{ab}$, where $a,b=\pm$. The indices $a$ and $b$ correspond respectively to the mixing and cubic vertex in Fig. \ref{fig:diagrams} (a). For example, $a=+(-)$ when the mixing vertex, comes from anti-time ordered (time ordered) part of the interaction Hamiltonian in eq. \eqref{ininmasterformula}. 

We will first evaluate $I_{-+}$ for which the time-ordered and anti-time ordered components factorize. We will do this in the squeezed limit where $k_1\approx k_2\gg k_3$ and denote the angle between $\vec{k}_1$ and $\vec{k}_3$ by $\theta$. 
\paragraph{Time ordered contribution.}
\begin{eqnarray}
(-i)\times \frac{2\psi_1(0)\dot{\phi}_0}{M_4}\times \int_{-\infty}^0 \frac{d\eta'}{\eta'^4}\eta'^2\times h_{k_3,0}(\eta)\times \frac{(1-ik_3\eta')}{2k_3^3}e^{ik_3\eta'}\nonumber \\=(-i)\frac{2\psi_1(0)\dot{\phi}_0}{M_4}\frac{m^2}{H^2}\times \frac{\mathcal{N}_{k_3}\sqrt{\pi}}{2\sqrt{2}k_3^2\cosh(\pi\mu)}.
\end{eqnarray}

\paragraph{Anti-time ordered contribution.}
\begin{multline}
(+i)\times \frac{\psi_1(0)}{M_4}\times \int_{-\infty}^0 \frac{d\eta}{\eta^4}\eta^4\times \epsilon_{ij}\bar{h}_{k_3,T}(\eta)\times (-i k_{1i})(-i k_{2j})\frac{(1+ik_1\eta)}{2k_1^3}\frac{(1+ik_2\eta)}{2k_2^3}e^{-ik_{12}\eta}\nonumber\\=
(+i)\times \frac{\psi_1(0)}{M_4}\times\frac{\mathcal{N}_{k_3}}{32k_1^4k_3}(\cos^2\theta-1/3)e^{i\pi/4}\times \nonumber\\
e^{-\pi\mu/2}\int_0^\infty dx x^{-\frac{1}{2}}\left[6x\left((2-i\mu)H_{i\mu-1}^{(1)}-(2+i\mu)H_{i\mu+1}^{(1)}\right)-(9-8x^2)H_{i\mu}\right](1-2ipx-p^2x^2)e^{2ipx}\nonumber\\
=(+i)\times \frac{\psi_1(0)}{M_4}\times\frac{\mathcal{N}_{k_3}}{32k_1^4k_3}(\cos^2\theta-1/3)e^{i\pi/4}\times (T_1+T_2+T_3),\nonumber\\
\end{multline}where
\begin{multline}
T_1=\left(8\mathcal{F}(\frac{3}{2},2p,\mu)-9\mathcal{F}(-\frac{1}{2},2p,\mu)+6(2-i\mu)\mathcal{F}(\frac{1}{2},2p,\mu+i)e^{i\pi/2}-6(2+i\mu)\mathcal{F}(\frac{1}{2},2p,\mu-i)e^{-i\pi/2}\right),\nonumber\\
T_2=-2ip\left(8\mathcal{F}(\frac{5}{2},2p,\mu)-9\mathcal{F}(\frac{1}{2},2p,\mu)+6(2-i\mu)\mathcal{F}(\frac{3}{2},2p,\mu+i)e^{i\pi/2}-6(2+i\mu)\mathcal{F}(\frac{3}{2},2p,\mu-i)e^{-i\pi/2}\right),\nonumber \\
T_3=-p^2\left(8\mathcal{F}(\frac{7}{2},2p,\mu)-9\mathcal{F}(\frac{3}{2},2p,\mu)+6(2-i\mu)\mathcal{F}(\frac{5}{2},2p,\mu+i)e^{i\pi/2}-6(2+i\mu)\mathcal{F}(\frac{5}{2},2p,\mu-i)e^{-i\pi/2}\right),
\end{multline}
and, 
\begin{multline}
\mathcal{F}(n,p,\mu) \equiv e^{-\pi\mu/2}\int_0^\infty dx x^n e^{ipx}H_{i\mu}^{(1)}(x)\nonumber \\
=(+i/2)^n\frac{1}{\sqrt{\pi}\Gamma(n+3/2)}\Gamma(n+1-i\mu)\Gamma(n+1+i\mu){}_2F_1(n+1-i\mu,n+1+i\mu,n+3/2,\frac{1-p}{2}).
\end{multline}
Using the asymptotic form of the hypergeometric function ${}_2F_1$ for large negative argument,
	\begin{equation}\label{hypergeolimit}
	{}_2F_1(a,b,c;z)=\frac{\Gamma(b-a)\Gamma(c)}{\Gamma(b)\Gamma(c-a)}(-z)^{-a}+\frac{\Gamma(c)\Gamma(a-b)}{\Gamma(a)\Gamma(c-b)}(-z)^{-b},
	\end{equation}
we can simplify the anti-time ordered contribution to get,
\begin{multline}
\text{Anti-time ordered contribution}=(+i)\times \frac{\psi_1(0)}{M_4}
\frac{3}{128 \sqrt{2} \pi}\frac{\mathcal{N}_{k_3}}{k_1^4 k_3}
 (\cos^2\theta - 1/3)\frac{1}{(1+4 \mu^2)}\times\nonumber\\
 \left(A(\mu)
     \left(\frac{k_3}{k_1}\right)^{1/2+i \mu} + A(-\mu) \left(\frac{k_3}{k_1}\right)^{1/2-i \mu}\right),
\end{multline}
where
\begin{equation}
A(\mu)=(-27 + 120 i \mu + 152 \mu^2 - 32 i \mu^3 + 16 \mu^4) \Gamma(5/2 + i \mu) \Gamma(-i \mu) 2^{-2 i \mu}.
\end{equation}
Multiplying the time and anti-time ordered contributions we get, 
\begin{multline}
I_{-+}=\frac{\psi_1(0)^2\dot{\phi}_0}{M_4^2}\frac{\sqrt{\pi}(\cos^2\theta-\frac{1}{3})}{128k_1^4k_3^2(1+4\mu^2)^2\cosh(\pi\mu)}
 \left(A(\mu)
     \left(\frac{k_3}{k_1}\right)^{1/2+i \mu} + A(-\mu) \left(\frac{k_3}{k_1}\right)^{1/2-i \mu}\right).
\end{multline}
Next we have to take into account $I_{+-},I_{++}$ and $I_{--}$. However, $I_{+-}$ and $I_{--}$ are just complex conjugates of $I_{-+}$ and $I_{++}$ respectively, hence we need only $I_{++}$. Computing $I_{++}$ analytically is difficult in general, however, in the squeezed limit $k_3\ll k_1$ we can get the non-analytic terms in $I_{--}$ by just making the variable change $k_1\rightarrow -k_1$ and changing the overall sign, i.e. for non-analytic pieces \cite{Arkani-Hamed:2015bza},
\begin{equation}
I_{++}(k_1,k_3) = -I_{-+}(-k_1,k_3).
\end{equation}
Using the above relation to sum over all diagrams and momenta gives finally (after reintroducing $H$),
\begin{multline}\label{f2}
F_{\text{KK Graviton}}^{\text{single}}=\frac{\psi_1(0)^2\dot{\phi}_0^2}{M_4^2H^2}\times (\cos^2\theta-\frac{1}{3})\frac{\sqrt{\pi}}{8 (1+4\mu^2)^2\cosh(\pi\mu)}\times  \\
\left(A(\mu)(1+i\sinh\pi\mu)
     \left(\frac{k_3}{k_1}\right)^{3/2+i \mu} + (\mu\rightarrow-\mu)\right).
\end{multline}
This can be equivalently written as,
\begin{multline}
F_{\text{KK Graviton}}^{\text{single}}=\frac{4\psi_1(0)^2\dot{\phi}_0^2}{M_4^2H^2}\times (\cos^2\theta-\frac{1}{3})\frac{\sqrt{\pi}}{ (1+4\mu^2)\cosh(\pi\mu)}\times  \\
\left(\frac{\frac{9}{2}+i\mu}{-\frac{1}{2}-i\mu} \Gamma(5/2 + i \mu) \Gamma(5/2-i\mu)\frac{\Gamma(-i \mu)}{\Gamma(1/2-i\mu)}(1+i\sinh\pi\mu)
\left(\frac{k_3}{4k_1}\right)^{3/2+i \mu} + (\mu\rightarrow-\mu)\right),
\end{multline}
whose form agrees with the results of \cite{Arkani-Hamed:2015bza,Arkani-Hamed:2018kmz} obtained via exploiting conformal symmetries of the late time slice.
\bibliographystyle{JHEP}

\bibliography{refs}

\end{document}